\RequirePackage{fix-cm}
\documentclass[smallextended]{svjour3}       
\usepackage{graphicx}
\usepackage{url}
\usepackage[breaklinks]{hyperref}

\usepackage{graphicx} 
\graphicspath{{figures/}}

\usepackage[numbers]{natbib} 

\usepackage[utf8]{inputenc} 

\begin{document}

\title{Machine Learning and Data Analytics for the IoT}


\author{Erwin Adi \and
        Adnan Anwar \and
        Zubair Baig \and
        Sherali Zeadally
}

\institute{E. Adi \at
	UNSW Canberra Cyber\\
	University of New South Wales, Australia\\
	\email{e.adi@unsw.edu.au}             \\
	\and
	A. Anwar, Z.Baig \at
	School of Information Technology\\
	Deakin University, Australia\\
	\email adnan.anwar@deakin.edu.au\\
	\email zubair.baig@deakin.edu.au
	\and
	S. Zeadally \at
	School of Information Science\\
	University of Kentucky, USA\\
	\email szeadally@uky.edu
}


\date{Received: date / Accepted: date}

\maketitle

\begin{abstract}
	The Internet of Things (IoT) applications has grown in exorbitant numbers, generating a large amount of data required for intelligent data processing. However, the varying IoT infrastructures (i.e. cloud, edge, fog) and the limitations of the IoT application layer protocols in transmitting/receiving messages become the barriers in creating intelligent IoT applications. These barriers prevent current intelligent IoT applications to adaptively learn from other IoT applications. In this paper, we critically review how IoT-generated data is processed for machine learning analysis, and highlight the current challenges in furthering intelligent solutions in the IoT environment. Furthermore, we propose a framework to enable IoT applications to adaptively learn from other IoT applications, and present a case study in how the framework can be applied to the real studies in the literature. Finally, we discuss the key factors that have an impact on future intelligent applications for the IoT.
	
\end{abstract}

\section{Introduction}

The Internet of Things (IoT) paradigm is both revolutionary as well as an enabler of automated and convenient lifestyles for modern day humans. The evolution of the IoT can be attributed to a confluence in advances that took place over the past decade in computing, communication and application design. The resulting sphere of influence of IoT has expanded rapidly to cover the whole human race. IoT devices in common use to facilitate our daily activities include the smart phones, home assistants such as Google Play, smart vehicles, building automation systems comprising smart elevators and temperature control systems, and unmanned aerial vehicles such as drones for environmental monitoring and leisure. The large-scale proliferation of IoT devices stretch beyond these devices to within the storage centers such as back end cloud facilities which are geographically dispersed. As a result, a large volume of data is generated by IoT devices and their supporting platforms, for transfer and subsequent storage and processing at back end cloud storage centers. IoT devices generate a constant stream of raw data, which cannot be discerned for meaningful knowledge unless the data is processed through application of techniques such as knowledge discovery and machine intelligence. The heterogeneity of the data generated from various IoT deployments is dependent on the application domain, comprising; smart healthcare, social media, e-agriculture, e-health, smart electricity grids and smart vehicles. IoT devices are designed with custom protocols that consider the resource constrained nature of these devices, in order to preserve power usage associated with device operations. The most common IoT application-layer protocols are Constrained Application Protocol (CoAP), Message Queuing Telemetry Transfer (MQTT), Advanced Message Queuing Protocol (AMQP), and HyperText Transfer Protocol (HTTP) \cite{EA018}. 

The MQTT and AMQP protocols are deployed on IoT devices with access to a continuous power supply, or to a renewable power source. Such protocols facilitate transfer of longer messages, and are thus more power-hungry. On the other hand, the CoAP protocol is light-weighted and is designed for highly resource-constrained IoT devices; where resource-constrained devices are those that have very limited computing resources and network bandwidth. HTTP protocol is the most resource-intensive IoT communication protocol, and is best suited for higher end IoT devices, possessing higher communication, computation and storage capabilities. IoT devices produce a large volume of data that is locally processed, in limited manner, and transferred to a centralized computing node or a cloud storage facility, where it can be further processed or analyzed to produce knowledge. 
Machine learning is defined as a family of techniques for analyzing data, wherein the process of model building on training data, is automated, i.e., requires little to no human intervention. Consequently, the process of categorizing data into various classes, is fully automated. The role of data analytics for IoT data processing cannot be understated, and machine learning is a very strong contributor to facilitate quick processing of large volume data emerging from IoT devices, for generating patterns of interest to analysts of the data. 

Contemporary computation and storage paradigms comprise; cloud, fog and edge computing. Through integration of these paradigms with the IoT, a robust data collection, storage, processing and analytics framework emerges. Such a framework has the ability to provide real-time insights into data patterns and also facilitates the application of machine learning techniques for realizing intelligent data analytics for the IoT. The cloud paradigm is a centralized model of data storage, that provisions various services such as Software as a Service (SaaS), Platform as a Service (PaaS) and Infrastructure as a Service (IaaS), to enable processing and analysis of IoT data, centrally \cite{J2}. Edge computing on the other hand enables the processing and analysis of IoT data closer to the IoT network, on localized computing nodes such as base stations. As a result, the cost associated with transfer of IoT data to centralized nodes, is avoided. Fog computing is a middle ground solution between Cloud and Edge computing paradigms, wherein the processing and analysis of IoT data need not be located at the edge of the network or even in a centralized storage facility. Rather, the fog paradigm brings forth the concept of a virtual platform, not exclusively located at the edge of the network, for processing and analysis of IoT data. 

In this paper, we study and analyze the role of machine learning to facilitate data analytics for the IoT paradigm. We present a thorough analysis of the integration of machine learning with the IoT paradigm in Section 2. In Section 3, we define the application of machine learning for processing and analysis of IoT data. We propose a novel framework for data analytics on IoT data emerging from various heterogeneous sources, in Section 4. In Section 5, we provide future directions of research work in the discipline of machine learning applications to IoT, and we finally conclude the paper in Section 6. 

\section{The convergence of machine learning and IoT}

The convergence of machine learning and IoT paves the way for a prospective advancement in efficiency, accuracy, productivity, and overall cost-savings for resource-constrained IoT devices. When machine learning algorithms and IoT work together, we can achieve improved performance for  communication and computation, better controllability, and improved decision making. Due to advanced monitoring from thousands to billions of ubiquitous sensing devices and improved communication capabilities, IoT has  enormous potential to improve the quality of human life and potential applications for industrial growth (toward Industry 4.0). IoT’s potential has significantly improved with the convergence of machine learning and Artificial Intelligence. Advanced machine intelligence techniques have made it possible to mine the huge volume of IoT sensory data to have better insights into a range of real-world problems, as well as the ability to make critical operational decisions. There- fore, to solve real-world complex problems and to meet the computation and communication requirements successfully, IoT and machine learning must complement each other. In recent years, IoT data analytics has gained significant importance and attention because of the following reasons:

\paragraph{High volume of data generated from distributed IoT devices:} 
According to the mobility report by Ericsson, the forecasting shows that there will be 18 billion connected IoT devices globally by 2022 \cite{EricReport}. This number will keep increasing over time due to the wide adoption of IoT devices in a wide range of critical applications. Intelligent data analytics will play an important role to identify and predict the future states of any process or system by mining this huge amount of data efficiently and intelligently.

\paragraph{High variability of data types from heterogeneous data sources:} 
Due to a wide range of applications and requirements, a large variety of IoT devices exist, which include mobile phones, PC/Laptop, tablets to short-range and wide- area IoT devices. Due to the heterogeneity of the data, the features, formats, and attributes of the data are different. Also, based on different IoT application domains, the data sources also vary. For example, the IoT devices used for medical applications will be different from a smart home IoT. Moreover, the quality, processing, and storage of data have also become a challenging task because of its heterogeneity. In~\cite{J8}, the authors highlight some key questions arising due to heterogeneity of data sources. It includes answers to critical questions such as: how to deal with the sampling procedure of the high-frequency streaming data, noise cancellation and filtering of the data, gathering and merging of the data from heterogeneous data sources, data interpretation and interoperability, reasoning, situation awareness and knowledge creation from the data, gathering and storing data from heterogeneous data sources to meet application’s constraints~\cite{J8}.

\paragraph{Uncertainty in the IoT data streams:} 
Uncertainty is very common in practical data analysis~\cite{Adnan_JCSS}. It may arise in the IoT data stream due to the failure of any IoT device or communication channel during data transfer. Gross errors and missing data are omnipresent in IoT data streams, which require advanced analytics to preprocess the data. Even cyber intrusion could be a valid reason for uncertainty in~data\cite{cyber_IoT}. In order to enhance the accuracy during decision making, it is critical to ensure the proper assessment, propagation, and representation of uncertainties and develop models and solutions that can deal with these factors~\cite{J8}.

\paragraph{Balancing scalability with the efficiency:} 
Most of the IoT data analytics are performed in the cloud. Transferring data from the IoT device to the cloud is expensive (in terms of delay), which may be challenging for time-critical applications especially when the number of IoT devices is high. For example, in a connected vehicle environment, a large pool of cars may be required to make decisions in real-time or near real-time. Here, it is important to balance the speed and accuracy of the analysis when the number of vehicles increases.

\subsection{Classification of various analytics techniques for IoT}

\begin{figure*}
	\includegraphics[width=0.8\linewidth]{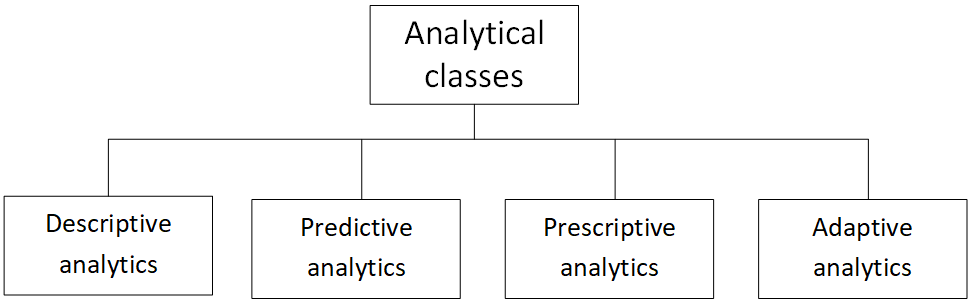}
	\caption{Major analytical classes}
	\label{fig:0anaclass}
\end{figure*}

In this section, we discuss different types of data analytics techniques. Figure~\ref{fig:0anaclass} shows the major component of the analytical classes.

\subsubsection{Descriptive Analytics}
IoT systems can gather data from a few to thousands of smart devices and transmit them to a cloud environment. Based on historical data, it is always possible to generate detailed insights into past events by using advanced ma- chine learning techniques. These groups of machine learning-based algorithms that process and summarize the raw data and provide actionable insights basically comprise the field of study called descriptive analytics. data Aggregation, data summarization, mathematical logical operations, data mining (e.g., clustering algorithms), and so on are some examples of descriptive analytics. Descriptive analytics requires a high-volume data. Recent technological advances have demonstrated that cloud storage is capable of storing huge volumes of IoT data, and cloud servers can process complex tasks using high-performance computers, and by applying  IoT cloud analytics.

\subsubsection{Predictive analytics for IoT}
Predictive analytics rely on historical data and utilize advanced statistical or machine learning techniques to model the behavior or pattern so that it is possible to predict the likelihood of possible future trends or patterns in data. To summarize, it predicts what will happen in the future by learning the historical patterns and data correlations of existing data. Predictive analytics have been widely used for different applications including predictive maintenance, prediction of price, supply-demand trend, or prediction of likelihood of any outcome. According to SAS, which is one of the top leading companies in analytics, there are two types of predictive models – (i) classification based models that conduct the prediction analyses by class membership, and (ii) regression-based models, that predict a number based on the historical observations and likelihood~\cite{Sas_ref}. State-of-the-art predictive modeling techniques include statistical regression-based models, Decision Trees, and Neural Network or Deep Neural Network-based models. Some other widely used algorithms are based on Bayesian analysis, Gradient boosting, Ensemble model-based analysis, and so on. These predictive analytics techniques are reliant on data for decision-making. The IoT paradigm can help facilitate the data gathering process from smart IoT devices and can provide an analytical framework using the cloud or the edge of the network.

\subsubsection{Prescriptive analytics for IoT}
Prescriptive analytics suggests how to respond to any future events based on data analysis. This class of analysis not only predicts the future states but also provides recommendations behind the adoption of the outcome. It is more like a future scenario analysis technique that leverages the benefits of both descriptive and predictive analytics. While predictive analytics suggests what and when the event will occur based on future predictions, prescriptive analytics extends the capability by providing insights of the future predictions with impact analyses. Prescriptive analytics is widely used to optimize the business outcome. Prescriptive analytics is suitable in an Industrial IoT (IIoT) setup where business intelligence-based decisions are made by using the capabilities of cloud/edge computing, big data analytics and machine learning. Services within an IoT-cloud platform can help to make optimal decisions through the deployment of business intelligence tools and through analytics.

\subsubsection{Adaptive analytics for IoT}
During actual implementation, the outcome obtained from the predictive analytics needs to adjust with real-time data. For this purpose, adaptive analytics are used to adjust or optimize the process outcome based on the recent history of the data and by looking at their correlations. This type of analysis helps to improve model performance and reduce errors. The advantage of adaptive analytics is that it can adjust the outcome of the solution when a new set of input data is received. Especially in an IoT environment, adaptive analytics are a good fit for real-time stream data processing. Real-time assessment of evolving data streams, such as those found in  malware \cite{ZB001}, can also be subject to adaptive analytics, to carry out data analytics.

\subsection{Classification of IoT data analytics based on Technological Infrastructure}

\subsubsection{Cloud Computing}
In a cloud computing paradigm, remotely located computing facilities (servers) are utilized using the Internet to store, gather, manage, and process the data. In the past decades, especially more recently, cloud computing has gained much attention because of the new infrastructure and processing architectures that it provides to support different services including Software as a Service (SaaS), Platform as a Service (PaaS), and Infrastructure as a Service (IaaS)~\cite{J9}. The cloud provides a variety of services delivered through the Internet, the concept of everything as a Service (XaaS) has emerged. In an IoT based environment, cloud computing can provide several advantages.

IoT interconnects smart devices. Within an IoT paradigm, a huge volume of generated data needs to be stored and analyzed. A cloud architecture may vary depending on the use case scenario, e.g., public, private, hybrid or community-based architecture~\cite{J9}.

\subsubsection{Edge Computing}
Edge computing simply refers to the analytical capability near the edge of the network of the IoT devices. The importance of edge computing can be better understood by realizing the challenges and limitations of cloud computing. Cloud computing is a centralized way of data processing and may increase the overhead of processing massive data \cite{J14}. Sending the raw IoT data streams to the cloud would increase the cost and will require high communication bandwidth and power. Instead, edge computing enables the computing within the ‘local’ network of the source of the data. Therefore, it solves the bandwidth problem of sending bulk amounts of data to the central server. Moreover, edge computing reduces the probability of a single point of failure and increases analytical efficiency. Compared with the huge and centralized servers as found in cloud computing, edge computing has smaller edge servers that are distributed. In \cite{J9}, it has been highlighted that cloud computing is suitable for delay-tolerant and complex data analysis, whereas edge computing is suitable for low-latency real-time operations. Besides, cloud computing requires fairly complex deployments when compared to edge computing. Fig.~\ref{fig:01} shows the architecture of a collaborative edge-cloud model for IoT networks. From the figure, the end-devices and the core network cloud are connected through an IoT gateway where the edge processing is performed. Typically, the deployment of edge computing requires minimal planning and in most cases can be ad hoc in nature \cite{J9}. As a result, it is predicted that around 45\% of the IoT data will use edge architectures in the near future \cite{EC_IoTJ}.

\subsubsection{Fog Computing}
Conceptually, fog computing lies somewhere in between cloud and edge computing and it acts as a bridge between the cloud and edge resources \cite{J2}.  The idea of fog computing was first proposed by Cisco \cite{Fog_cisco}, to address the issues associated with latency-sensitive applications. According to \cite{Fog_cisco}, unlike edge computing where computing happens at the edge of the network, fog computing provides networking services between the cloud and end-devices as well as provides computing and storage facilities through a virtual platform.
As highlighted in the Cisco whitepaper \cite{Cisco_whitepaper}, the fog nodes have time-sensitive data analytics capabilities and are a good fit for applications or services with response times ranging from milliseconds to minutes. However, the data storing capabilities of the fog nodes are very limited. On the other hand, cloud technologies can store data for up to months or even years \cite{Cisco_whitepaper}. 

\begin{figure*}
	\includegraphics[width=\linewidth]{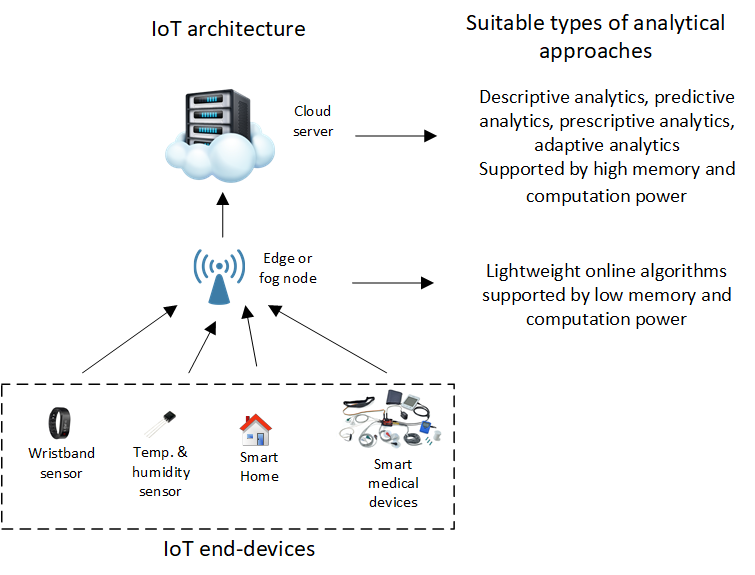}
	\caption{Convergence of IoT architecture and data analytics: an illustration}
	\label{fig:ana_iot}
\end{figure*}

\begin{figure*}
	\includegraphics[width=\linewidth]{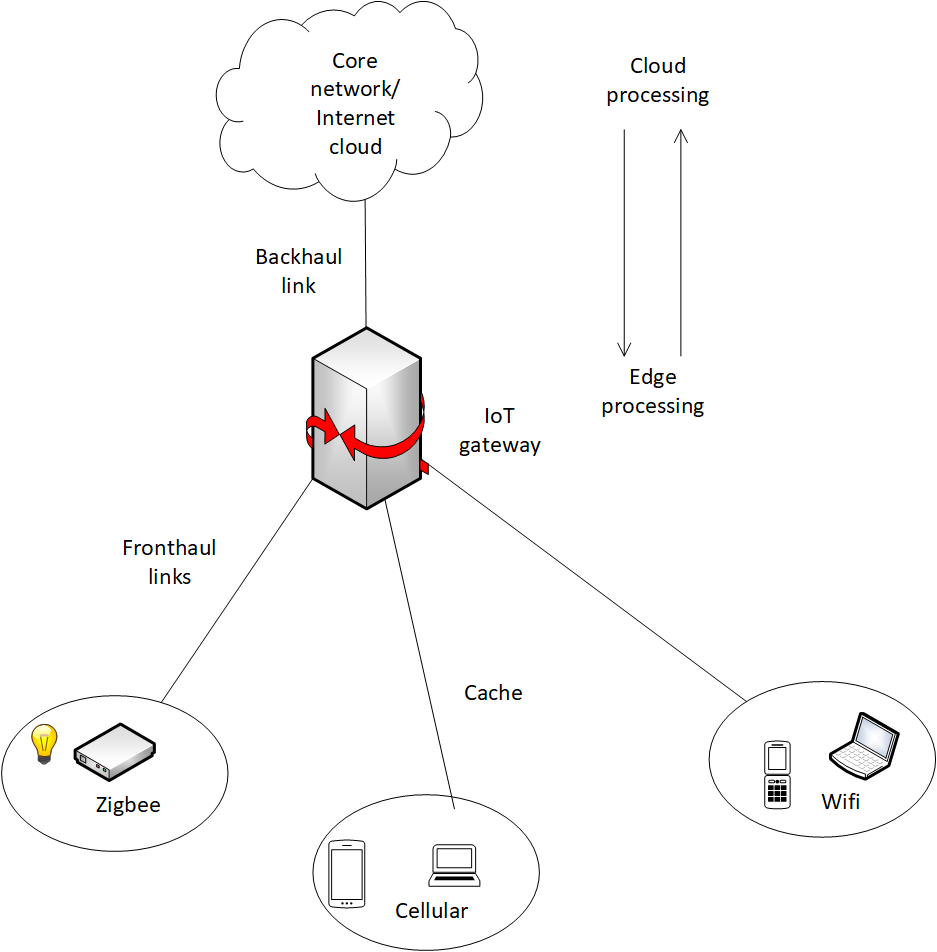}
	\caption{A generalized system model for collaborative edge-cloud processing in heterogeneous IoT networks~\cite{J9}}
	\label{fig:01}
\end{figure*}

Fig.~\ref{fig:ana_iot} shows various component of IoT and how analytics relate with these components. Because of the high computational resources and processing capabilities, most of the analytical approaches are suitable in the cloud server of an IoT setup. On the other hand, lighweight and online algorithms are better fit for the edge computing.

\subsection{Applications of IoT Data Analytics}

\subsubsection{IoT data analytics for smart vehicles}

IoT has enormous potential in the connected vehicle environment, especially for efficient and accurate decision making using advanced data analytics. Within an Internet of vehicle (IoV) paradigm~\cite{Sherali7892008}, a large number of broadcasting messages are frequently generated with very high granularity and volume. Therefore one of the biggest challenges is storing and intelligent management of the huge amount of data \cite{V1}. Another key issue is related to ensuring the security of the data, as any cyber-related anomalies or cyber intrusion attempts will jeopardize the system and may cause fatalities. Hence, it is important to look for intelligent solutions that can deal with such cyber-related incidents. Moreover, a centralized solution approach may not be feasible as it is more prone to a single point of failure. In \cite{V1}, the authors utilize a blockchain-based approach for distributed and secure storage of vehicular data.  The authors also define how the nodes in the connected vehicle environment could be integrated into blockchains through a blockchain-based network architecture. One significant difference between connected vehicular networks and other IoT application areas is that vehicles have high-speed mobility. Due to the environmental impact and the nature of the open wireless medium, high-speed mobility could be a reason for vehicle data faults \cite{V2}. This area requires significant research attention. Haibin et al. proposed the threshold-based fault detection and repairing scheme using a dynamic Bayesian network (DBN) model in \cite{V2}. The model considers both spatial and temporal correlation of the connected vehicle data. This work shows that their proposed analytical framework can detect faults with good accuracy and a low false alarm rate. 

He et al. proposed two advanced analytical frameworks for vehicle warranty analysis in the connected vehicle environment \cite{V3}. Two improved data mining algorithms based on the Bayes model and a Logistic Regression model, are proposed. These machine learning algorithms are used for effective data mining service development using cloud data analytics. The authors also demonstrate the intelligent parking cloud service \cite{V3}. Typically, it is very challenging to find an available parking lot, especially in urban area. Moreover, it may lead to traffic congestion, stress and increase the probability of accidents. To solve this challenge, a `birth-death' model is proposed in \cite{V3}. The stochastic `birth-death' is a special case of continuous-time Markov process. This model is utilized in order to predict the occupancy probability of any parking space.   

An autonomous and real-time method has been proposed for crime detection on public bus services in \cite{V4}. The authors highlight the advantages of IoT technologies and demonstrate the advantage of data analytics within a fog computing environment. A framework has been developed for intelligent public safety in a connected vehicle environment. Advanced data analytical capability for detection, prediction, and prevention of crime has been the key focus of this study. The proposed framework shows significant improvement in performance and device survivability over the traditional smart transportation safety use cases \cite{V4}

\begin{figure*}
	\includegraphics[width=\linewidth]{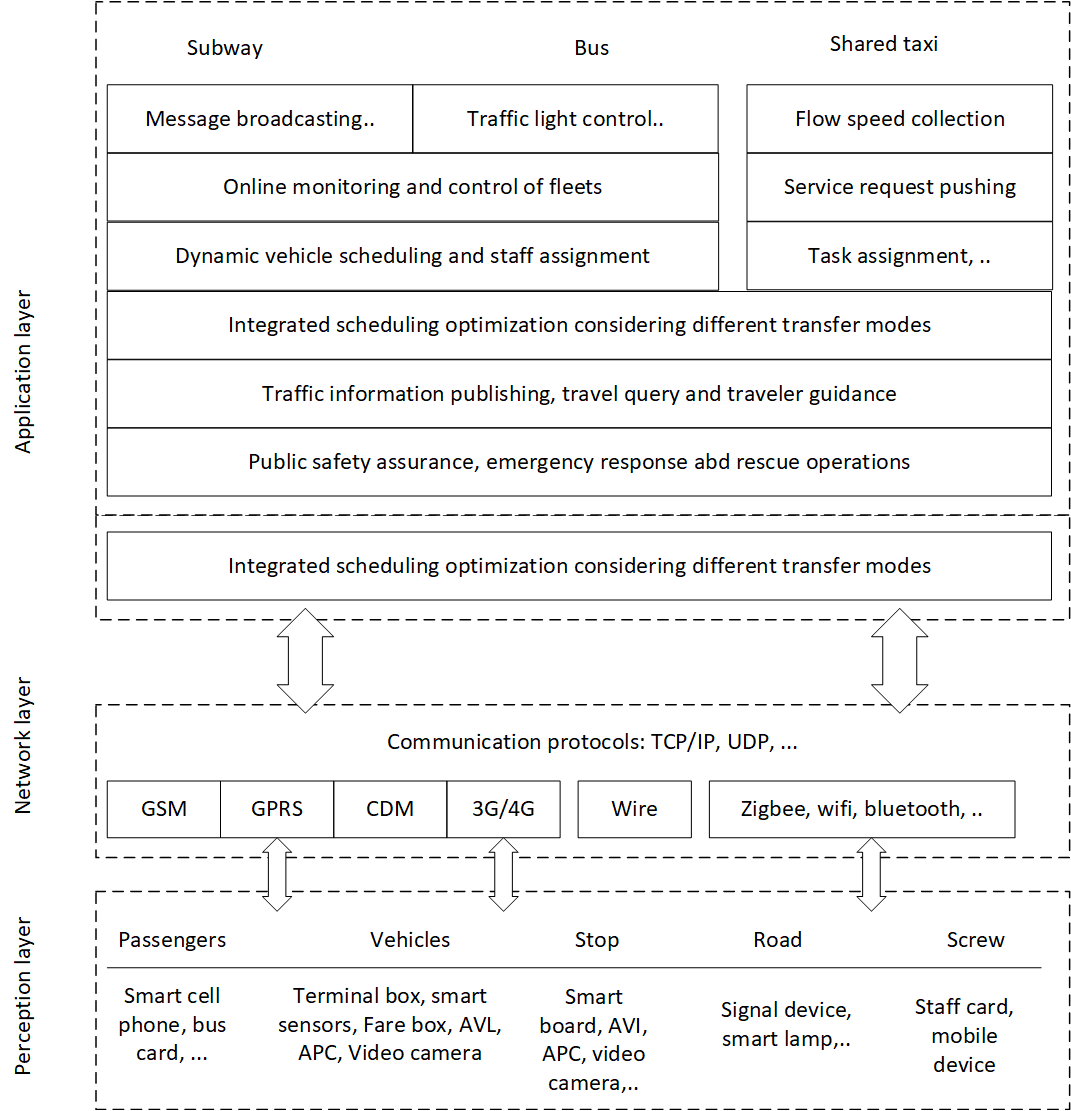}
	\caption{A three-tier architecture of an intelligent transportation system based on IoT~\cite{V5}.}
	\label{fig:02}
\end{figure*}

Other than crime detection, IoT connected vehicular data analytics can be of use for other critical applications. For example, Luo et al proposed a three-tier framework of the connected public transportation system for effective traffic management \cite{V5}. Here, the authors propose innovative solutions for effective scheduling of subway, bus, and shared taxis in an IoT connected environment. A periodic pattern mining based algorithm is deployed for determining the passenger and road flows. Finally, the capability of evolutionary-based algorithms is used to develop a computational model for dynamic bus scheduling \cite{V5}. Fig. \ref{fig:02} shows a three-tier architecture of the IoT based system for several applications in a connected vehicular environment using data analysis and effective communication. In the first layer (Perception layer), data is gathered from various IoT sensors and devices. For example, a passenger’s smart cell phone can provide the locational information of the passenger, whereas a smart travel card may contain information related to the passenger’s identification and financial transactions \cite{V5}. Other smart IoT devices of the environment are smart terminal boards, Automatic Vehicle Location (AVL), Automatic Passengers Counting (APC), digital video cameras, and devices installed for road safety like smart road cameras, etc. \cite{V5}. The second layer (Network layer) in the framework is responsible for transferring the source data to the application layer (final layer) using both wired and wireless communication. Standard communication technologies are used, e.g., Mobile Communication (GSM), General Packet Radio Service (GPRS), Code Division Multiple Access (CDMA), the 3rd Generation communication (3G), and the 4th Generation communication (4G), etc. for public wireless communication and Zigbee, Wi-Fi and Bluetooth for private wireless communication \cite{V5}. In the Application layer, IoT data is used by advanced analytics. Here, optimization and machine learning algorithms are used for decision making of various applications deployed for subway, bus and shared taxi networks.   

Intelligent parking space allocation has been the focus of research presented in \cite{V6}. Kong et al. develop a systematic approach using IoT and cloud computing to solve the parking allocation problem. The scheme utilizes auction-based mechanisms to solve the parking allocation problem. In \cite{V7}, Priyashman et al. explore the capability of two machine learning algorithms for vehicular access control and identification. Identification of two key critical factors – signal strength and tag settings- has been the prime focus of their work. Authors in \cite{V7} show that machine learning algorithms like linear regression and logistics regression analysis can provide an acceptable outcome to identify these critical parameters.

Swarm intelligence-based decision-making approach has been proposed in \cite{V8}. Considering an IoT connected vehicle environment, this work shows that smart vehicles can make dynamic and adaptive decisions. Here, the authors show how IoT based communication is effectively utilized for traffic intensity calculations and subsequent utilization of IoT data for dynamic decision making using swarm intelligence \cite{V8}.

\subsubsection{IoT data analytics for smart healthcare}

\begin{figure*}
	\includegraphics[width=\linewidth]{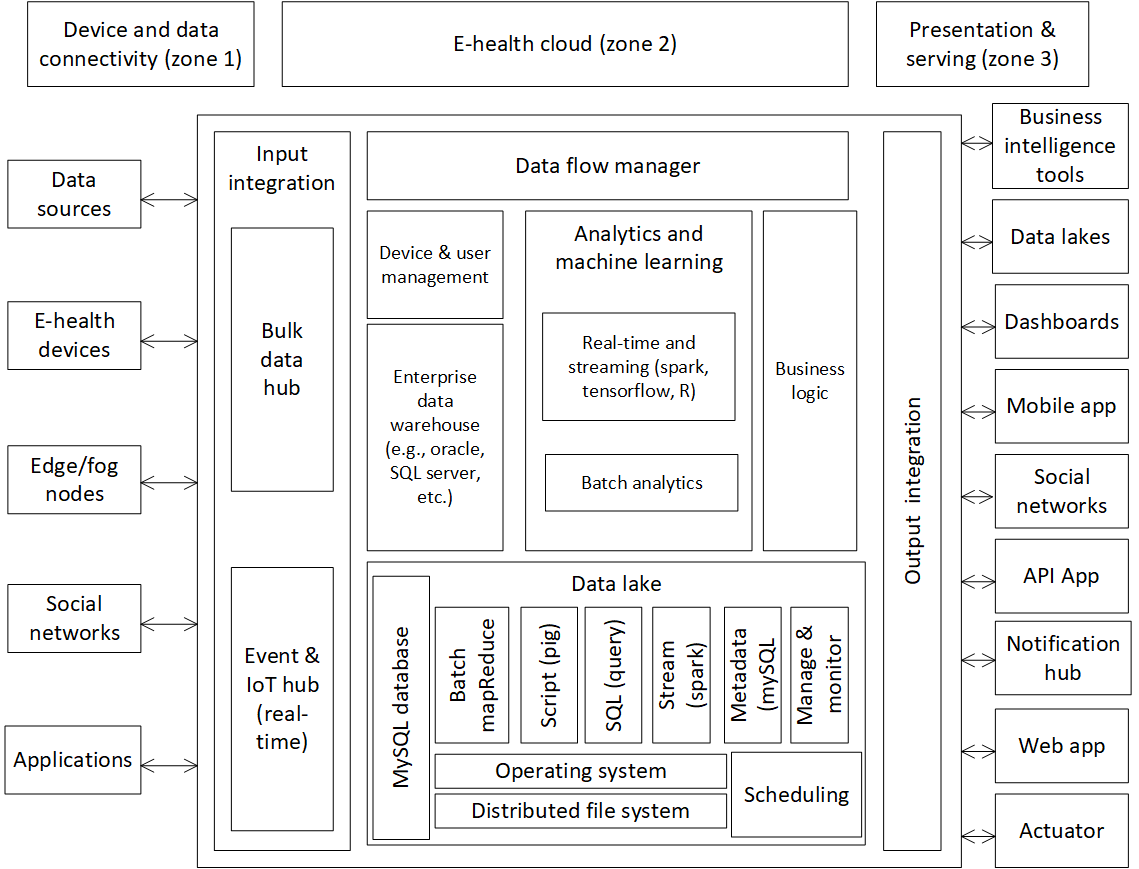}
	\caption{Multilayer architecture of the eHealth cloud~\cite{HC_2}.}
	\label{fig:03}
\end{figure*}

The rapid growth of IoT has benefited different application areas including healthcare~\cite{ZEADALLY2019100074}. IoT technology along with data analytics can help to achieve more accurate and improved health diagnoses~\cite{HC_2} .   
One critical requirement is to gather data, predict and make decisions in real-time. In \cite{HC_2}, the authors highlight the importance of the real-time pattern recognition technique deployment for the construction of genomics-based patient models. A dynamic and adaptive computational model with intelligence needs to be developed. These innovative models should be able to capture the data produced by thousands of IoT connected nodes within the smart healthcare paradigm. Advanced analytics and communication within a connected healthcare system can ensure lots of benefits including – improved resiliency, seamless fusion with different technologies, big data processing and analytics, personalized forecasting of patient condition, lifetime monitoring of patient health, ease of use of wearable devices, overall medical health cost reduction, physician oversight with real-time patient data, availability and accessibility of the doctors through advanced communication and efficient healthcare management \cite{HC_2}. It is also highlighted that the model-based approach, which is widely used in the industry, may not be a suitable fit for the health domain because medical data changes continuously and is highly prone to uncertainty. Therefore, instead of a model-based approach, a data-driven approach could be a better solution \cite{HC_2}. Fig. \ref{fig:03} shows a multilayer architecture of the eHealth cloud~\cite{HC_2}. While the data generation and communication takes place in Zone 1 (see Fig.~\ref{fig:03}), the main computational process is conducted in Zone 2; real-time and streaming analytics are located here. Tools such as Hive, Spark~\cite{dash2019big,Ed-daoudy2019}, MapReduce~\cite{Ed-daoudy2019}, HDFS, YARN are used for this purpose in Zone 2~\cite{HC_2}. Finally, outcomes are displayed or actuated based on the requirement of the application or services in Zone 2.

A smart sleep monitoring system is proposed in \cite{HC_1}. The authors propose a fog computing-based preprocessing approach implemented on the smart device. Next, the capability of big data on the cloud is utilized by performing batch data processing. Predictive analysis is performed to identify the air quality for the treatment of Obtrusive sleep apnoea (OSA) \cite{HC_1}.  The outcome of the analysis is displayed using web user interface to benefit the health professionals in real-time irrespective of their location of operation.

An IoT enabled automated nutrition monitoring system is proposed in \cite{HC_3}. The authors propose a meal-prediction algorithm using a detailed analysis of Bayesian networks. To investigate the nutrition balance, another algorithm is proposed that uses perceptron neural network-based deep learning models. This computational model utilizes the smart IoT framework for monitoring and communication. 

Early detection of Dyslexia, which is a cognitive disability that affects the regular activities of an individual, is possible by using a ledger-based technology within an IoT framework \cite{HC_4}. The proposed approach gathers data from smart IoT devices such as smartphones, tabs or laptops, and stores the data using blockchain-based distributed ledger technology. Due to the use of blockchains, it is available to the medical practitioner for evaluation irrespective of the location, and it also provides security of the information.

An agent-based IoT simulation testbed has been developed in \cite{HC_6} to investigate a patient’s sleeping behavior. In the context of a smart bed, the simulator analyzes the sensor information to detect the posture during sleep time. An emotion detection model is proposed in \cite{HC_7}. The proposed solution receives the speech and image signals from IoT devices. These signals are analyzed on the edge cloud and remote cloud to make the decisions. Fourier transform, different filtering algorithms and Support Vector Machines (SVMs) are used for the detection. The proposed framework is suitable to deploy in a big data environment.

\begin{figure*}
	\includegraphics[width=\linewidth]{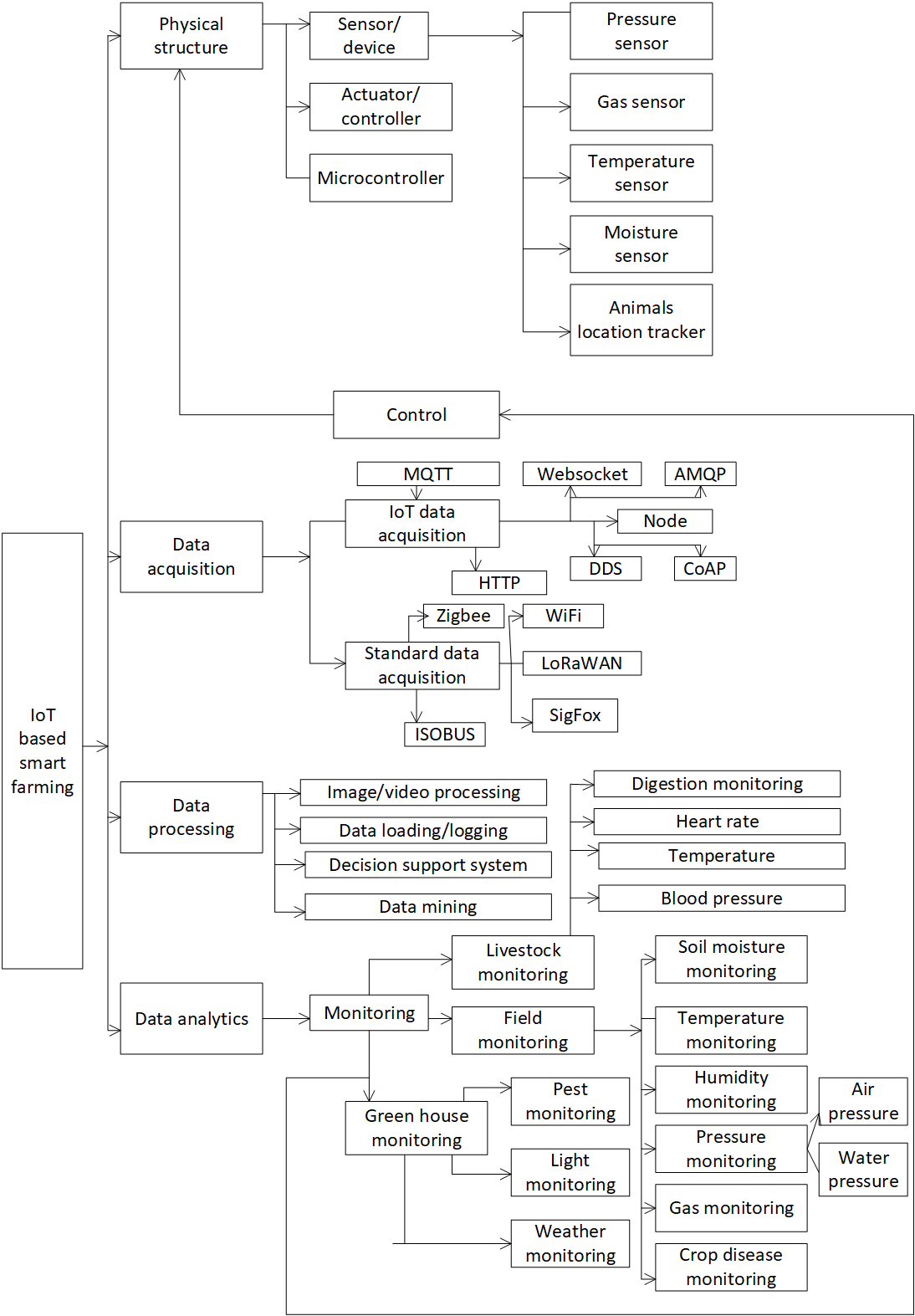}
	\caption{Major components of IoT-based smart farming \cite{Agri2}}
	\label{fig:04}
\end{figure*}

\subsubsection{IoT data analytics in Agriculture}

IoT based frameworks can be adopted to improve the operational efficiency and accuracy by using advanced analytics of the data generated from smart end devices. Several challenges and benefits of IoT and data analytics-enabled smart agriculture frameworks are discussed in \cite{Agri1}. In the recent past, the prospects, opportunities, and feasibility of Wireless Sensor Networks (WSN) in agriculture have been widely studied \cite{agri_external2}. The application of WSN in agriculture includes environmental monitoring, precision, agriculture, machine and process control automation and traceability \cite{agri_external1, Agri1}.  Recently the importance of IoT has been highlighted because of its versatility of the adoption of different wired and wireless technologies as well as the capability to integrate with advanced data analysis mechanisms. IoT in agriculture empowers the farmers with advanced and improved automation and decision making processes~\cite{Agri1}. It also enables seamlessly integration among agricultural products, services and knowledge to improve the quality and productivity~\cite{Agri1}.

IoT is deployed for agricultural asset monitoring and management. RFID based tags are also used for tracking and tracing of supply chain products \cite{SheraliRFID}. GPS can also be used for the same purpose within an IoT framework along with edge/cloud data analytics~\cite{Agri1}. Typically, the humidity, wind speed, temperature, cloud transients, rainfall and weather parameters are very critical for effective farming. IoT with machine intelligence can also help to predict the localized weather of the agricultural firms. Moreover, the efficiency of agricultural waste and water management can be improved with the smart use of IoT and the big data paradigm. By using intelligent scheduling algorithms, the challenge of storage management can be resolved. IoT devices can be placed strategically to monitor the storage facilities and then cloud technologies can deploy machine learning and optimization algorithms to make the decisions \cite{Agri1}. 

Fig. \ref{fig:04} shows the major component of IoT based smart farming. Out of the four major components, three are related to the data (highlighted in the shaded box). The physical structure controls the sensors, actuators and devices and is responsible for the overall precision in data processing \cite{Agri2}. Data acquisition and data processing are similar to those found in standard IoT applications, based upon state-of-the-art IoT protocols. For the data analytics conponent, various algorithms and models are used for data analytics. Some IoT and data analytics based smart agriculture applications include livestock health monitoring, stress level monitoring, physical gesture recognition, Rumination, heart rate tracking, livestock location monitoring, climate condition monitoring, pest identification, irrigation management, greenhouse gas monitoring, and so on~\cite{Agri2}.

\subsubsection{IoT data analytics in Energy Systems}
In recent years, the energy grid has transformed significantly due to the integration of photovoltaics solar cells, electric vehicles, and storage with the low-voltage distribution network. These distributed energy resources need to be effectively coordinated and controlled. One viable solution is to use the capabilities of smart meter. Within an IoT environment, smart meters are interconnected and linked with each other. Advanced analytical decisions are necessary for improving the efficiency and reliability of energy operations. Therefore, in \cite{J18}, the authors pointed out that the combination of IoT and Big Data can be used for effective energy management.
The authors develop an energy management system that can be used in a smart home. Based on the unique ID address for each IoT object, a System-On-Chip (SoC) module is deployed for data acquisition. A centralized server is used within a Home Area Network, where data is collected, stored and processed. One advantage is that the proposed solution is capable of using off-the-shelf business intelligence tools. The solution architecture can significantly improve the efficiency of high energy consumption applications such as air conditioning. This has been validated in a lab environment. Considering a smart home architecture, IoT and sensor data optimization based peer-to-peer energy trading architecture is proposed in \cite{Sebastian8909771,electronics8080898}

Complex Event Processing (CEP) is used for reactive applications to provide a distributed but scalable solution within an IoT environment \cite{J31}. The combination of CEP and historical data has been used to develop a prediction algorithm using machine learning algorithms in \cite{J31}. Here, the authors extend the work for a dynamic IoT data environment by proposing an adaptive predictive model using the moving window regression technique. To determine the optimal window size, authors use a spectral component analysis of the time series data.

In \cite{SG2}, the authors surveyed the possibilities and opportunities of big data and cloud analytics for the smart grid. Once data is collected from the smart meters, tools in a cloud platform can perform advanced analytics. IBM Coremetrics~\cite{EA301} and Google BigQuery~\cite{EA302} are some examples of cloud-based software solutions for data analytics.

Predictive analysis for decision support systems of a smart meter IoT network has been presented in \cite{SG3}. Here, the authors use a machine learning-based approach for cost prediction and to improve the performance of the network operation. In this work, a Bayesian network model is compared against three other classifiers including Naïve Bayes, Random Forest and Decision Trees. The proposed method in \cite{SG3} is validated using the network coverage data collected from commercial settings. 

The IoT connected smart grid integrated with machine learning can provide several benefits including predictions of consumption, price, power generation, future optimum schedule, fault detection, adaptive control, sizing, and intrusion or anomaly detection \cite{SG4}. For such applications, connectivity and data exchange plays a significant role. IoT is being widely deployed for this purpose. In \cite{SG4}, the authors also discussed the security issues in IoT data analytics. Lots of research~\cite{AA8010846,AA2015201,SNI8777171,GUNDUZ2020107094} has been done to investigate security issues of traditional energy grid and Supervisory Control And Data Acquisition (SCADA) systems using various attack detection algorithms~\cite{WANG201942,He7926429}, there is a need to analyze the impact of cyber security on IoT connected smart grid. In \cite{SG4}, the authors also highlighted that machine learning can play an important role in cyber-attack detection in smart grid.     

A machine learning-based multi-stage theft detection system is proposed in \cite{SG5}. 
In the first stage, multi-modal forecasting is used for predicting the consumption behavior . Multi-Layer Perceptron (MLP), Recurrent Neural Network (RNN), Long Short Term Memory (LSTM), and Gated Recurrent Unit (GRU) based machine learning models are used in this stage.  
In the second stage, decisions are made. Here, a moving average model is used to identify the anomalies. In the final stage, the historical energy usages are used to decide whether the anomaly based on stage 2 is an energy theft or not~\cite{SG5}. 

A lightweight machine learning-based Intrusion Detection System (IDS) for IoT is developed in \cite{SG8}. The proposed IDS is capable of detecting Denial of Service (DoS) attacks successfully. The authors use SVM based classification techniques to develop the IDS. Unlike other works, the proposed work considers only one attribute to model the classifier. This assumption made the model computationally efficient and helped achieve faster solutions. The comparative results presented in this paper show that the SVM-based IDS outperforms the state-of-the-art techniques in accuracy.

\section{A critical review on data processing and knowledge discovery for IoT}

\subsection{Transforming data to knowledge}
\begin{figure*}
	\centering
	\includegraphics[scale=0.6]{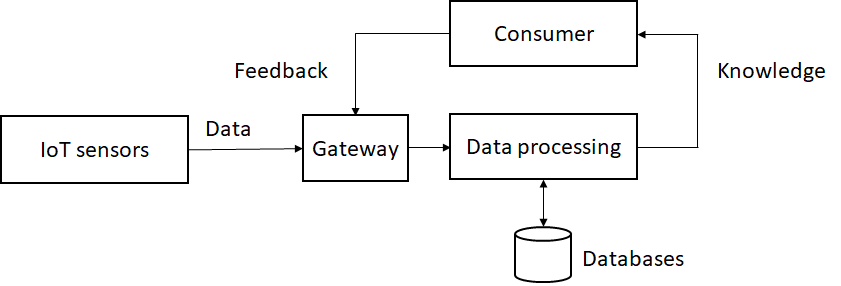}
	\caption{The architecture of knowledge creation in the IoT, adapted from \cite{EA001}.}
	\label{fig:EAF01}
\end{figure*}

IoT devices must process data to produce meaningful knowledge. How data becomes meaningful to its consumer is illustrated through the architecture shown in Figure \ref{fig:EAF01}. In this architecture, the consumer can be a machine or human. When the consumers expect to receive meaningful information, the data processing unit consists of AI algorithms such as machine learning techniques. In this article, meaningful processed data is also referred as "knowledge". 

Prescriptive, predictive, and adaptive learning in the IoT can be best described through how the applications process data to create knowledge. The above architecture is used throughout this section to organize the evaluation on how current studies implement AI in the IoT paradigm. This section highlights the implementation challenges, discusses how studies address them, and identify future research opportunities.

\subsection{The application-layer protocol}
In machine-to-machine communications, an application-layer protocol is required to send data in the format that can be encapsulated by the lower layer protocols. As mentioned previously, the widely used IoT application-layer protocols are Constrained Application Protocol (CoAP), Message Queuing Telemetry Transfer (MQTT), Advanced Message Queuing Protocol (AMQP), and HyperText Transfer Protocol (HTTP). All these protocols allow for abstraction from the lower layers.

\begin{figure*}
	\centering
	\includegraphics[scale=0.6]{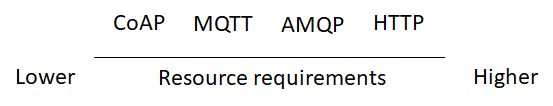}
	\caption{IoT application protocols, i.e. Constrained Application Protocol (CoAP), Message Queuing Telemetry Transfer (MQTT), Advanced Message Queuing Protocol (AMQP), and HyperText Transfer Protocol (HTTP). Adapted from \cite{EA018}. }
	\label{fig:EAF05}
\end{figure*}

Constrained devices have limited computing, communication, and storage resources. Since different IoT devices have varying resource requirements, the protocols differ in their capabilities to meet these requirements. Resource requirements were derived from the message sizes, overhead, and latency. As illustrated in Figure \ref{fig:EAF05}, CoAP is the most lightweight, while HTTP is the most resource intensive. CoAP is used in applications where IoT data is collected from constrained devices at the edge. These devices are typically light in weight and they include wearable devices or the remote, low maintenance sensors. They have very limited computing resources, power and capability to transmit data wirelessly. Such applications therefore do not send knowledge back to the IoT devices. Hence, devices employing CoAP are suitable for prescriptive and predictive AI applications, but not adaptive.

MQTT and AMQP are employed when the applications can afford higher-end IoT devices. The IoT devices are accessible for power recharge or are directly connected to the power supply. In these applications, data can be forwarded to other machines or consumers across the Internet, including back to the IoT devices. Hence, this scenario can be used for prescriptive, predictive, and adaptive AI applications.

\begin{figure*}
	\centering
	\includegraphics[scale=0.6]{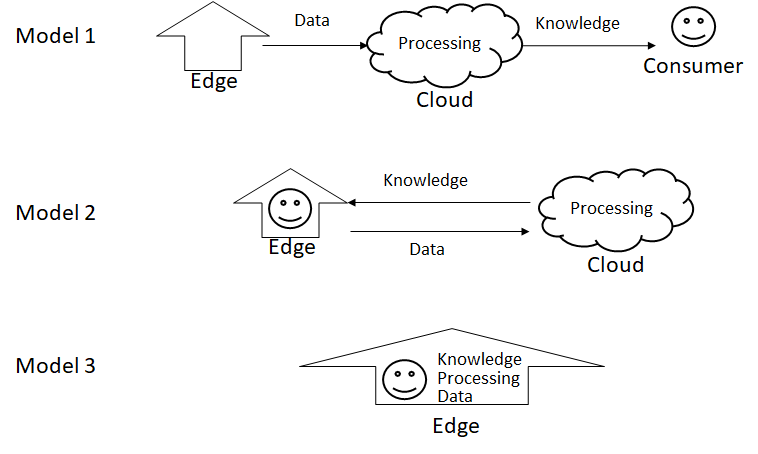}
	\caption{Where data is transformed to knowledge.}
	\label{fig:EAF02}
\end{figure*}

On one hand, the application protocols have addressed how constrained devices can send data. On the other hand, machine learning analysis require high computing resources and a large amount of data to process the collected data to convert it to meaningful knowledge. Therefore, IoT applications vary the architecture of knowledge creation (Figure \ref{fig:EAF01}) to transform data to knowledge based on the framework and the network bandwidth available for data transmission. Figure \ref{fig:EAF02} illustrates three models on how the architecture of knowledge discovery can be applied.

The first model is when a constrained device sends data to a remote system in a simplex manner. This is the case in data mining, where the data is used or shared with different applications. Each remote application can have its own method to transform data to knowledge. Adaptive learning takes place in the cloud, since feedbacks cannot be sent back to the IoT device. In other words, the IoT device is unable to adaptively learn.

In the second model constrained devices can send/receive data in duplex communications. The remote system, e.g. the cloud, processes the data and brings knowledge back to the device. This model can be seen in many smartphone applications. Adaptive learning may be processed either on the data processor in the cloud or on the IoT device.

In the third model,  smart devices have enough computing resources to turn data to knowledge locally or at the edge. This is suitable for delay-sensitive applications. Adaptive learning is therefore processed at the edge, either on the data processing machine or on the IoT device.

Table \ref{tab:EAT01} illustrates how application protocols were employed in the current literature. The table shows the applications of the studies and the type of data collected from the sensors to illustrate the suitability of protocols to transfer the data. The table also shows the type of model (according to Figure \ref{fig:EAF01}) that the studies adopt. The table presents the knowledge to illustrate the results from the data processing done by the machine learning techniques adopted in the studies. The three models show that, although application protocols address how IoT applications can adopt AI to some extent, there are some limitations on how IoT devices adaptively learn.

\begin{table}
	\caption{Knowledge created from the data produced by different applications.}
	\label{tab:EAT01}       
	\begin{tabular}{|p{36mm}|p{13mm}|l|p{23mm}|p{15mm}|p{18mm}|}
		\hline
		\noalign{\smallskip}
		\textbf{Application (from the title of the publication)}                                                                                                     & \textbf{Protocol}         & \textbf{Model} & \textbf{IoT data}                                                                                                             & \textbf{Machine Learning} & \textbf{Knowledge}                                                                                   \\
		\noalign{\smallskip}\hline\noalign{\smallskip}

		Predicting Energy Consumption Through Machine Learning Using a Smart-Metering Architecture \cite{EA033} & MQTT, CoAP       & 1     & Energy consumption, temperature, humidity, precipitation, wind speed, time direction; all at 15-min interval         
		& Random forest
		& Predicted energy consumption                                                                \\
		\hline
		
		Smart City Traffic Monitoring System Based on 5G Cellular Network, RFID and Machine Learning \cite{EA141}                                                & MQTT             & 1     & Magnetic, light, temperature, fuel consumption, time to reach destination, vehicle weight, road slope                
		& Azure machine learning
		& Optimum traffic route                                                                       \\
		\hline
		
		Namatad: Inferring occupancy from building sensors using machine learning \cite{EA142}                                                                   & MQTT             & 1     & Temperature, CO2, air volume, HVAC temperature                                                                & Random forest
		& Predicted occupancy level                                                                   \\
		\hline
		
		Lameness Detection as a Service: Application of Machine Learning to an Internet of Cattle \cite{EA131}                                                   & MQTT             & 1     & Step count, lying time, swaps per hour                                                                               
		& Random forest, k-NN
		& Lameness level (active, normal, dormant)                                                    \\
		\hline
		
		Case study: Integrating IoT, streaming analytics and machine learning to improve intelligent diabetes management system \cite{EA134}                     & AMQP             & 1     & Blood sugar reading                                                                   
		& Azure machine learning
		& Predicted sugar level                                                                       \\
		\hline
		
		An IoT system to estimate personal thermal comfort \cite{EA132}                                                                                          & AMQP             & 1     & Heart rate, skin temperature, room temperature, humidity, air speed                                                  
		& Support vector machines
		& Thermal comfort level                                                                       \\
		\hline
		
		Cloud computing based smart garbage monitoring system \cite{EA135}                                                                                       & MQTT             & 1     & Level of garbage in a bin, timestamp                                                                                 
		& Decision forest regression
		& Predicted bin filling pattern                                                               \\
		\hline
		
		An IoT-Based Solution for Intelligent Farming \cite{EA029}                                                                                               & MQTT, AMQP       & 1     & Sheep posture: neck inclination and distance to the ground                                                           
		& Rule engine
		& Feed on vines or weeds                                                                      \\
		\hline
		
		IoT and distributed machine learning powered optimal state recommender solution \cite{EA133}                                                             & MQTT             & 2     & Location, acceleration, heart beat                                                                                   
		& Kalman filter
		& Product recommendation                                                                      \\
		\hline
		
		Quantifying colorimetric tests using a smartphone app based on machine learning classifiers \cite{EA136}                                                 & AMQP             & 2     & Average red, green, blue values                                                                                      
		& Support vector machines
		& Quantified peroxide content                                                                 \\
		\hline
		
		On Delay-Sensitive Healthcare Data Analytics at the Network Edge Based on Deep Learning \cite{EA137}                                                     & CoAP             & 3     & Pulse rate, respiratory rate, oxygen level, sleep condition, fall detection, gait tracking, washroom visit frequency 
		& Deep learning
		& Anomalous health risk                                                                       \\
		\hline
		
		Fall detection system for elderly people using IoT and ensemble machine learning algorithm  \cite{EA021}                                                 & CoAP, MQTT, HTTP & 3     & Accelerometer data, gyroscope data                                                                                   
		& Ensemble random forest
		& Fall (forward, backward, lateral) and normal activities (walking, stairs climbing, sitting) \\
		\hline
		
		Vibration Condition Monitoring Using Machine Learning \cite{EA138}                                                                                       & CoAP             & 3     & Motor vibration level in mV                                                                                          
		& Neural network
		& Motor condition (normal, unbalanced)                                                        \\
		\hline
		
		Early Detection System for Gas Leakage and Fire in Smart Home Using Machine Learning \cite{EA139}                                                        & MQTT             & 3     & Temperature, humidity, gas, smoke, CO, flame, CO2                                                                    
		& Classification and regression trees
		& Disaster risk levels                                                                        \\
				
		\hline                                                                                                                               
	\end{tabular}
\end{table}

A recent study \cite{EA021} showed how a combination of these application-layer protocols can be employed to exchange knowledge in the IoT environment. Different protocols were selected according to their degree of constraint. CoAP was used to allow local communication between a wearable device and an IoT gateway; HTTP was used to exchange knowledge between the IoT gateway and the cloud solution; and MQTT was used to send alert from the IoT gateway to the subscribers such as emergency services and family members. As Table \ref{tab:EAT01} shows, combining protocols are commonly adopted to support the varying resource requirements for different types of applications. 

Current studies employing IoT data for machine learning focus on how the consumers can benefit from the IoT technologies by proposing where the data should be processed, and how an application layer protocol selection can be justified to fit its purpose. Literature reviews \cite{EA018, EA019} compared how the application-layer protocols exchange data based on their client-server architecture, resource discovery framework such as Representational State Transfer (REST) and Service-oriented Architecture (SOA), the transport protocol such as Transport Control Protocol (TCP) and User Datagram Protocol (UDP), data query method such as Structured Query Language (SQL) and publish/subscribe, security protocol, and packet header/message size. However, few studies have compared how the current application layer protocols adaptively learn knowledge in machine-to-machine communications. Although Table \ref{tab:EAT01} shows that knowledge is created in these studies, machine-to-machine autonomous learning was not fully explored. Creating knowledge from data is considered as the function of a multipurpose box wherein all required data is assumed to have been available before data processing employing AI or machine learning techniques.

\subsection{The gateway}

Simply put, a gateway is a multipurpose box at the edge to perform the necessary action \cite{EA028}. A gateway converts, routes, or even processes data. It may be used as a filter that selects data of interest and drops the noise \cite{EA009, EA028}, or as hub that aggregates data from several IoT devices \cite{EA002}. While aggregated data at the edge may be too big to be transmitted over the Internet, gateways can be designed as the data processor to produce and transmit low bandwidth messages \cite{EA002, EA021}. Depending on the system specification, a gateway may not be needed, or may be integrated with the devices.

\begin{figure*}
	\centering
	\includegraphics[scale=0.6]{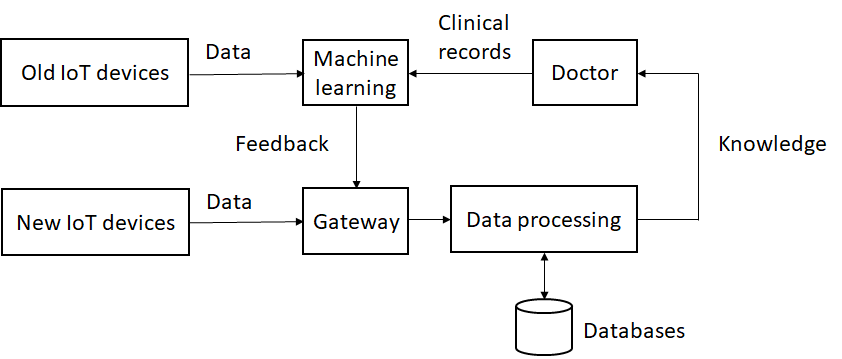}
	\caption{Gateways adapt heterogeneous data from different devices.}
	\label{fig:EAF06}
\end{figure*}

The problem in processing data from the IoT sensors is due to the heterogeneity of the data. Consider a health monitoring system that employs wearable devices to generate data for medical devices \cite{EA004}. Data heterogeneity issue will arise when the number of new wearable IoT sensors grows at a much faster pace than how the medical devices could adapt to the variety of the generated data. The data is unlabeled and varies in dimension. Therefore, preprocessing the data is needed for its use in supervised machine learning techniques. In this study \cite{EA004}, data from the old IoT devices was adapted to handle the data from the new devices through a gateway (Figure \ref{fig:EAF06}). Clinical records from doctors were used to train machine learning techniques, to produce data that can synchronize the old and the new wearable devices. Here, the gateway serves as a data adaptor.

Data from IoT devices may be represented in raw format. Hence, gateways are devised to transform raw data to a usable format, such as in the form of features that can be employed for machine learning analysis \cite{EA004, EA023, EA021}. The transformed data may be encapsulated in a network packet with a header and other meta-information such as sensor messages and addresses, for transmission to other devices \cite{EA024}.

As such, gateways can act as an adaptor that convert one protocol to another. At the application-layer, they adapt based on the protocol used at the edge to the one used by the cloud. For example, from constrained IoT devices employing CoAP to a cloud service that communicates through MQTT or AMQP \cite{EA024, EA021, EA027}. They may also convert the lower-layer wireless protocols of the local, constrained devices for enabling Internet based transmission to the cloud \cite{EA033}. At the network layer, gateways may be used to optimize routing, for example, by selectively prioritizing packets in networks connecting vehicles, which are bursty in nature. In this case, a gateway is deployed to minimize the number of delayed packets \cite{EA025}.

Since gateways perform many tasks, it is challenging to determine the computing resources required to deploy one. Therefore, studies show how their resource consumption can be measured \cite{EA022}. Relevant evaluation for deployment includes how throughput and latency perform under varying loads, such as increasing the  of clients when running a combination of services.

Despite having gateways perform the necessary actions, they were designed for and evaluated under each specific case study only. Current studies in employing IoT data for machine learning emphasize on how to adapt data heterogeneity from varying sensors, pre-process the data and synchronize communications with the cloud. However, the solution was not directly adopted to adapt heterogeneous data for different studies. In other words, current challenges include not only adapting data from new devices, but also from new knowledge, i.e., meaningful data generated from AI machines. While a gateway is a device to perform the necessary action, there is a need to integrate it with the necessary actions required for discovering knowledge.

\subsection{Deriving knowledge from data}
\label{sec:cep}
\begin{figure*}
	\centering
	\includegraphics[scale=0.6]{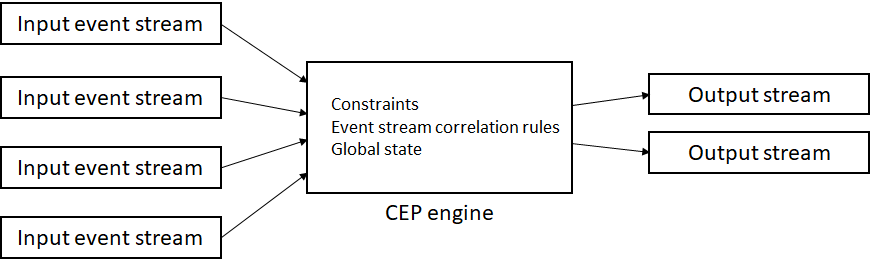}
	\caption{Complex Event Processing (CEP) engines listen to events from input streams to process data, adapted from \cite{EA040}.}
	\label{fig:EAF10}
\end{figure*}

IoT sensors generate a stream of raw data, much of which is meaningless to the consumer. Raw data is a blend of insight and noises. It does not show  a change in events, quantization per time, or meaningful attributes and relationships used by the consumers. The challenge to derive meaningful data from sensors is apparent in the IoT when the pervasive sensor devices deliver varying context and media. In this case, data processing in the IoT follows the method described in the Complex Event Processing (CEP) approach, which takes events from input streams to process data, regardless of the technology or protocol employed (Figure \ref{fig:EAF10}).

Capturing events implies creating processable data out of the raw data. Data is sent as input streams, events are extracted from some patterns such as a change of state or time in the data. A series of states can create a pattern of interest. However, meaningful patterns can become stale when a system fails to process them within a specific time. Thus, systems that derive meaning from input patterns implement event listeners.

Event listeners are implemented in many programming languages, with Java presenting an elegant implementation for parallel processing; and as one that can integrate low-level events into management-level events \cite{EA040}. This management-level information is the knowledge gained. Hence, in CEP, knowledge is implemented as maps between event layers. For example, a global pandemic watch system (as knowledge) was implemented as correlations between various event layers such as news feeds around the globe, Short Message Service (SMS) messages from trained agents in some designated area, and electronic reports from health authorities \cite{EA040}.

When correlating events, a knowledgebase is needed to define what the correlations are. While the relationship definitions are traditionally provided by experts, advancements in machine learning techniques allow self-assigned rule creation from input streams. In this case, the knowledgebase is created as rules generated from a set of supervised rule-learning algorithms \cite{EA046}. The problem with supervised machine learning techniques is that the labeling process in creating training datasets still requires expert interventions rather than reliance on machine-to-machine communications.

Rather than manual intervention, ontologies have been used to address how machines label data. An ontology is a network of objects describing their properties and relationships. Given an ontology of a domain, labels can reveal their meaning within the domain. One method of describing meaning is by showing the object property in a label-value pair. For example, in the domain of Vehicles, events such as speed can be labeled with low, medium, and high-speed, each with a range of values. A recent study showed how machines can self-assign their labels, given speed values at the input event stream \cite{EA045}. Another method to describe meaning is by expressing the object relationship in a subject-predicate-object form. For example, given a knowledgebase with an entry ”access point X is in location Y” and ”A has phone B”, one may derive knowledge that ”A is in location Y” upon an event ”B is connected to access point X” \cite{EA044}. In both above scenarios, knowledge is derived through machine-to-machine communications.

While ontologies address how data can be labeled and associated, current work still requires a holistic system that can share outputs between different machines \cite{EA040}. However, current reviews on CEP are concerned with how the approach is implemented and how the various implementations affect the overall throughput, CPU time cost, and communications cost \cite{EA042}. Consequently, more recent studies have proposed different optimization strategies such as query reordering, memory management and parallelization, with each strategy compared in terms of time complexities. Studies in CEP work are concerned with CEP performance rather than how machines can exchange meaningful data.

\subsection{Machine-to-machine knowledge exchange}
The Semantic Web was proposed to explain how machines can semantically communicate with other machines \cite{EA006}. One fundamental component in Semantic Web is the ontologies concept. As previously described, an ontology is a taxonomy that defines a set of objects and their relationships. It can therefore represent a set of inference rules in AI.

\begin{figure*}
	\centering
	\includegraphics[scale=0.6]{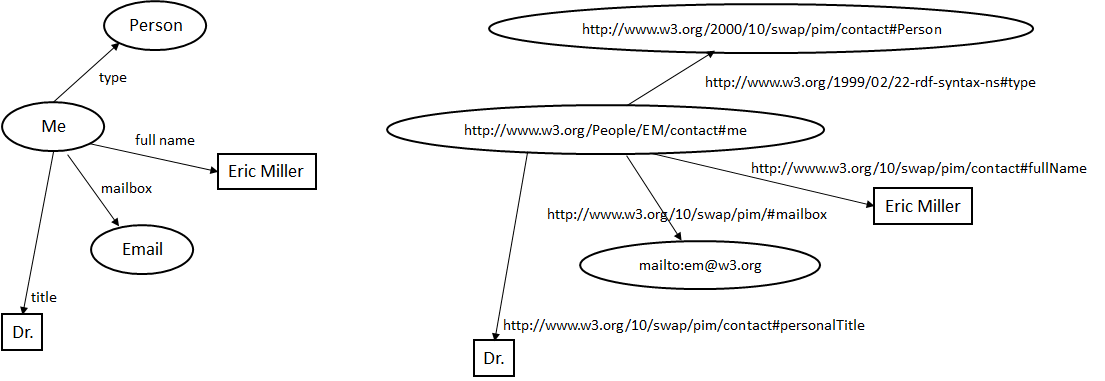}
	\caption{Ontologies, adapted from \cite{EA007}. (a) The ontology of a Person. (b) Employing an application-layer protocol to describe an ontology. }
	\label{fig:EAF03}
\end{figure*}

Figure \ref{fig:EAF03}a shows an ontology of a Person in terms of a graph (adapted from \cite{EA007}). The nodes represent objects, and the edges represent relationships. A member of a Person object is Me, which has a name, email, and a title attribute. These relationships can derive an inference rule ”if an object has an email, it may be a person.”

The Semantic Web employs HTTP to exchange the information describing graphs. This is illustrated in Figure \ref{fig:EAF03}b. Each node and edge have a unique Uniform Resource Identifier (URI), which is manifested in the form of a web address. Hence, the graph allows machines to follow the ”scent” to find information about a specific object, such as who Eric Miller is and what his email address is. This demonstrates how an application-layer protocol, HTTP, provides a method for knowledge exchange in a machine-to-machine communication. Locating object properties and relationships is not confined through URI as described by the Semantic Web. Competing approaches include service-oriented knowledge discovery, which will be demonstrated in the next section.

In its deployment, a gateway can be installed as a hub at the edge of network to exchange ontology information with the cloud using HTTP, while facilitating local communications with the IoT devices under CoAP \cite{EA021}. The deployment can be an implementation of a study to alert remote emergency services when elderly people are in need, such as when a fall is detected.  The method used to recognize a ”fall” requires knowledge about how an anomalous event can be detected by device sensors. In this study, knowledge was exchanged between the gateway and a remote database to describe what normal/anomalous daily activity features are. The database is equipped with an ontology that ”a Device is carried by a User” and that ”a Device has Events”. Upon receiving a meaningful input event, the gateway (rather than the IoT devices) can distinguish anomalous from normal daily events.

The above study demonstrates that the problem with adopting Semantic Web in the IoT is that HTTP is not suitable for constrained devices (as was previously shown in Figure \ref{fig:EAF06}). HTTP requires higher computing and network resources compared to other IoT application-layer protocols such as CoAP. The header and message size are limited to a maximum of 2 bytes and 256 MB respectively in CoAP, compared to the unspecified length in HTTP \cite{EA018}. These fundamental differences affect the way they can be adopted in an IoT environment. That is, adopting Semantic Web to exchange knowledge degrades the system performance when CoAP is employed. Therefore, the current challenge in machine-to-machine knowledge sharing is how constrained IoT devices can exchange semantic messages and can globally find the ontology that the messages belong to. The literature has compared how machines exchange IoT data but not knowledge.  Current work in intelligent IoT applications requires a review not only in communication protocols and their performance, but also in how physical/digital modules can smartly communicate.

\subsection{Smart objects}
Physical or software objects are regarded as smart when they autonomously exchange knowledge to solve problems. Such smart objects are aware of and respond to the changes in their environment and recognize some opportunities to proactively make a decision. They must show sociability with other smart objects or people to accomplish a goal, and support other applications  \cite{EA089}.

Most smart object implementations make use of ontologies, or some ap- proach that involves a knowledgebase system consisting of objects, their properties and relationships  \cite{EA083}. For example, in a task analysis domain, such a knowledgebase can consist of a hierarchy of task decomposition starting from the goal, a breakdown of the state changes, to the low-level task definitions \cite{EA084}. 

Many implementations have local rather than shared knowledgebases. An example of a rich knowledgebase digital system is a Building Information Model (BIM), which is a digital platform originally employed for managing building constructions for planning and cost effectiveness. The model breaks down the information of a building, from room layouts to construction materials required, forming a hierarchy of properties and relationships. With the introduction of IoT, BIM applications advanced to build green buildings by predicting and simulating energy models \cite{EA201}; and promote safety by detecting hazardous gases on the location where the building workers were situated \cite{EA202}. Despite having state-of-the-art tools for designing knowledgebases, BIM applications do not have a universally defined ontology to eliminate data interoperability issues \cite{EA200}.

\begin{figure*}
	\centering
	\includegraphics[scale=0.6]{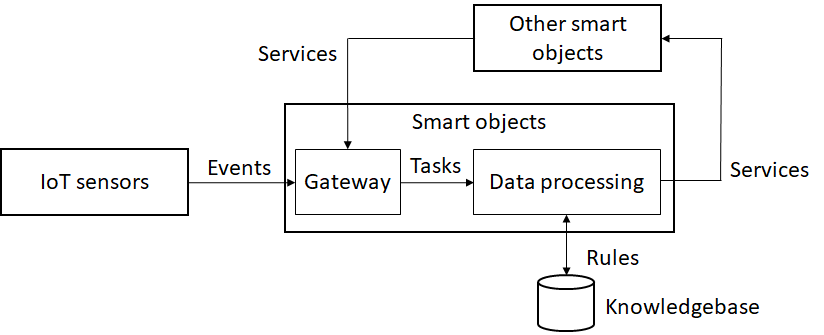}
	\caption{Smart Objects share their services, enabling them to learn the breakdown of rules to a set of tasks and events.}
	\label{fig:EAF09}
\end{figure*}

Challenges in implementing smart objects include how to discover and locate shared knowledgebases. Hence, design frameworks proposed in the literature present a discovery module to find how a system relates to other smart object systems and their ontologies \cite{EA083}. Such a module is analogous to a gateway. In discovering how other systems accomplish a goal, the module allows smart objects to hierarchically break down how a task is described by a knowledgebase \cite{EA081}, where a task consists of a range of events, and relationships express inference rules. This is illustrated in Figure \ref{fig:EAF09}. The decision from processing these events and rules represents the services that a smart object can advertise to other objects. Newly discovered services can be updated in the knowledgebase, extending the ontology, and in return creating new knowledge. 

Despite smart objects having shown the ability to exchange and share knowledge, the use of the constructed ontology is not widely deployed. Current work is focused on the development and evaluation of ontologies rather than the usage of the constructed ontologies \cite{EA203}. Although ontologies allow objects to define and share properties and relationships of a domain, the ability to semantically identify services disintegrates when the ontology is used in other domains. In the various BIM ontologies for example, different situations cause dividing interpretations into whether certain data is mandatory or optional  \cite{EA204}. In addition, rule inference tasks have shown to be difficult to adapt given the varying BIM ontologies. A survey indicated that two-third of the ontologies reviewed were not reused, while there was no indication whether the remaining one-third reused ontologies could be applied to a different domain \cite{EA205}.

\subsection{Summary}
Message exchange between IoT devices is constrained by their computing power and network bandwidth. On the other hand, smart applications require big data and massive computing resources for machine learning-based analysis. Therefore, current smart applications adapt to such limitations by proposing machine learning analysis at the gateway or in the cloud. These applications generate prescriptive and predictive analysis; but implementing adaptive applications has shown to be challenging in a constrained computing environment. Adaptive applications require feedback from either human or other smart machines, which require a duplex communication channel. In addition, feedbacks are processed by algorithms that are evolutionary, which require more resources than what can be achieved by constrained devices. Hence, IoT devices are limited in their ability to autonomously learn, build, and share knowledge. 

To address its limitations, IoT applications employ ontologies, i.e. databases describing a network of objects through their properties and relationships. This allows IoT solutions to locate inference rules or knowledge that is built in other machines or the cloud, creating intelligent systems for an application. Despite ontologies being widely applied to build smart applications, adopting one ontology to a different domain has shown to be difficult. The ability of an IoT application to semantically identify the meaning of messages weakens when applying an ontology from a different domain.

\section{A framework for machine learning and knowledge discovery for IoT}
Despite the pervasiveness of IoT devices, the inadequacy of machines to adaptively learn becomes the barrier in the proliferation of intelligent IoT systems. Little has been studied in how machines can autonomously share knowledge when the data come from different domains or case studies. Thus, this section proposes a framework on how machines can autonomously exchange knowledge, create new knowledge, and adaptively learn from the knowledge so that they can become applicable across different domains or case studies. The framework is adopted from how humans learn and classify knowledge. To begin with, this section reviews the various case studies associated with intelligent big data analytics for IoT.

\subsection{A review of intelligent big data analytics for IoT}
Table \ref{tab:EAT01} shows the various case studies that have leveraged big data analytics to address IoT challenges by employing machine learning techniques. The studies can be grouped into four domains: smart city, production (i.e. farming \& manufacturing), building management, and health. The following discussions describes the studies given in Table \ref{tab:EAT01}.

The smart city domain is concerned with bringing efficient and enjoyable daily living activities. Some examples include providing optimum traffic route \cite{EA141}, predicting garbage bin filling pattern for collection \cite{EA135}, recommending a product based on one's location \cite{EA133}, and predicting energy consumption in smart meters \cite{EA033}. In predicting energy \cite{EA033}, Support Vector Machines were used to analyze the past energy consumption and the environment data of a building such as the temperature and humidity \cite{EA033}, resulting in a predicted energy consumption with 1.7 kWh difference between the actual and predicted consumption in a 30-day window.

In the production-farming domain, in \cite{EA131}, the authors proposed to detect lameness symptoms in dairy cows. Diseased cows affect milk productions, and their symptoms are usually displayed through inactive behaviors such as lying down for a duration of time. Thus, IoT sensors were placed on cows to collect their activity data such as lying time, swaps per hour, and step count. Machine learning techniques revealed three kinds of behaviors such as active, normal and dormant in cows. Random Forest gave an accuracy of 91 percent, detected 1 day before some visual signs can be observed; and k-Nearest Neighbors gave an accuracy of 81 percent, detected 3 days before their symptoms were noticeable through the visual signs. 

In the production-manufacturing domain, the maintenance of machinery is essential to minimize production interruptions due to failed motors. Thus, a study \cite{EA138} placed 3-axes accelerometers on factory motors to collect vibration data such as the amplitude and frequency of the motors. With Neural Networks, the study was able to identify faults from normal motor conditions with 100 percent accuracy, with confidence level between 80 percent and 99 percent. 

Building management may be seen as a domain that connects the smart city and manufacturing domain, because their solutions may be applied to both domains. A study \cite{EA142} predicted the occupancy level of a building by observing the temperature, CO2, air volume, and air conditioning data. Random Forest was employed to classify the data and yielded 95 percent accuracy in predicting the occupancy level of rooms within the building. Another study \cite{EA139} collected building sensor data such as temperature, humidity, gas, smoke, CO, flame and CO2 to equip the building with a early detection system against gas leakage and fire hazard. The study arranged four risk levels, from no risk to high risk; and the regression trees algorithm achieved 99.93 percent accuracy in predicting the risk levels.

The health domain may be the most pervasive applications because the IoT devices are attached to individuals. The applications predict sugar level for diabetes management \cite{EA134}, estimate thermal comfort level at a workplace \cite{EA132}, and detect when a person falls at home \cite{EA021}. A study \cite{EA137} employed deep learning to detect ambulation events such as abnormal walking pattern, sleeping habits, and washroom visits. The study provides a real-time solution to detect anomalous health risks from wearable device data collecting heart rate, respiration rate, and so on, and yielded a 94 percent accuracy. Such a solution may be viewed as a smart home application because they also bring efficiencies in daily activities (e.g. providing a cyber guardian for elderly people).

\subsection{Learning from human-computer interaction}
How a human can interact with websites can teach how machines should interact with other machines. The most relevant property of a website is that its presentation can be understood by humans. Users can semantically navigate websites to accomplish the goals required of a task. Websites consist of words in the language that is understood by the human users. In addition, website layouts are arranged in a meaningful structure exhibiting their "information scent" so that the users can naturally navigate to find the information being searched. 

There are two characteristics of knowledge sharing that can be drawn from human-computer interaction. Thus, we propose  the following theorems.

\begin{theorem}
	Naming an entity (i.e. object, properties, relationship and services) must use words that humans understand. 
\end{theorem}

\begin{theorem}
	Information must be arranged semantically in a way that it allows human to derive meaning by following its structure. 
\end{theorem}

Naming an entity must use words in human language rather than abbreviation, codes, or binary mapping that only machines can interpret, despite the latter being technically more efficient in terms of storage space or network bandwidth.

In arranging information semantically, the use of ontologies is the model itself. It allows humans to retrieve information based on semantic queries. In this work, we propose the use of ontologies for machines to autonomously share knowledge and adaptively learn from other machines. How this can be articulated may be derived from how human classify knowledge.

\subsection{Epistemology}
The philosophy of knowledge, or epistemology, provides the breakdown of how a human gains knowledge. Fundamental in this area is "a priori" and "a posteriori" knowledge. A priori is gained through the definitions, such as the classification; apples are fruits. A posteriori is gained through experience and observations, such as apples are red.
 
Variations from the above definition are also used to describe ideas and human understanding. This is illustrated in Table \ref{tab:EAT02}. Kant’s \cite{EA110} a priori and a posteriori description of knowledge are parallel to Locke’s \cite{EA111} primary and secondary quality in human understanding. Primary quality includes the properties of an object independent of an observer, e.g. an apple has weight, size and color. Secondary quality refers to some object properties according to an observer, e.g. apples are red. Descartes \cite{EA112} used an additional term, i.e. invented, which denotes ideas gained from imagination, such as the mermaids and the unicorns.

\begin{table}
	\caption{The categories of knowledge.}
	\label{tab:EAT02}       
	\begin{tabular}{llll}
		\hline
		\noalign{\smallskip}
		Knowledge gained through:            & Kant \cite{EA110} & Locke \cite{EA111} & Descartes \cite{EA112} \\
		\noalign{\smallskip}\hline\noalign{\smallskip}
		Definition & A priori     & Primary       & Innate            \\
		Experience                                                  & A posteriori & Secondary     & Adventitious      \\
		Imagination                                                 & -            & -             & Invented          \\
		\noalign{\smallskip}\hline                                  &              &
	\end{tabular}
\end{table}

Henceforth, in this work we adopt the word \textit{Primary}, \textit{Secondary}, and \textit{Invented} to classify knowledge.

\begin{itemize}
	\item \textit{Primary} knowledge is built into ontologies, such as the definition of "If the soil is wet then crops flourish".
	\item \textit{Secondary} knowledge is gained from the IoT sensors, such as "The soil is dry".
	\item \textit{Invented} knowledge is created after abduction inferences, speculating what else serve as the indicators to have the crops flourished such as the sunlight and fertilizers.
\end{itemize}

Secondary knowledge can become Primary knowledge when the quality it describes can be generalized. This agrees with Berkeley \cite{EA114} who contemplated that primary knowledge comes from something that is perceived. Similarly, when Invented knowledge (such as a hypothesis) has been scientifically proven, the parameters it describes become Secondary knowledge. This agrees with the verificationism theory \cite{EA113}, which serves as the groundwork for the proposed framework of machine knowledge in our paper explained below.

\subsection{A framework of machine knowledge}
\begin{figure*}
	\centering
	\includegraphics[scale=0.6]{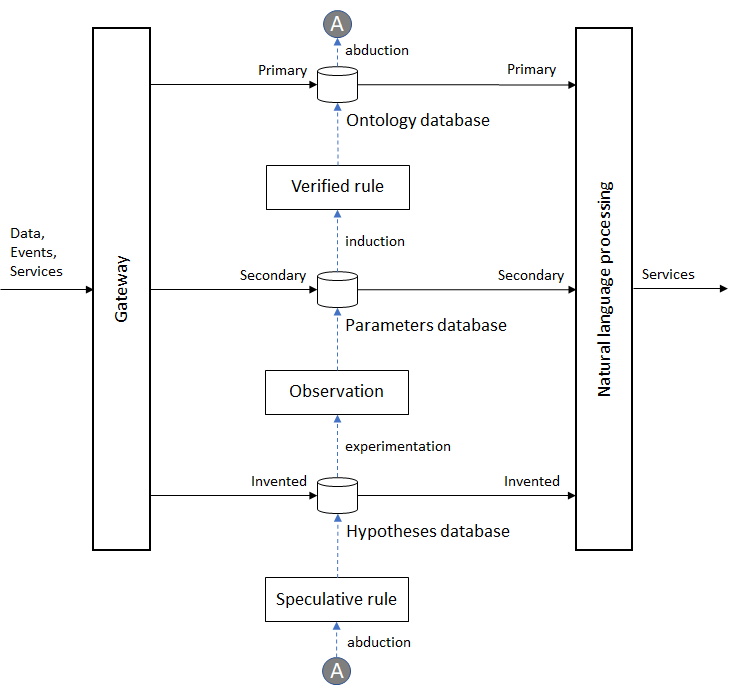}
	\caption{Smart Objects share their services, enabling them to learn the breakdown of rules to a set of tasks and events.}
	\label{fig:EAF11}
\end{figure*}

The framework is adapted from the verificationism theory \cite{EA113} which provides a breakdown of scientific methods. It is a school of thought where knowledge is gained from experimentally verified observations. As the society has developed through this scientific pathway, its framework in gaining knowledge through observations can be borrowed for the design of autonomous learning machines. The framework of machine knowledge proposed in our paper is shown in Figure \ref{fig:EAF11}.

The framework describes how Smart Objects (SOs) can autonomously exchange knowledge from other SOs. It consists of three databases, namely the Ontology, Parameters, and Hypotheses database. When sending data from the databases, SOs tag the services to signify the Primary, Secondary, or Invented level of knowledge, accordingly. SOs exchange knowledge by advertising and searching for services with other SOs as what previously was shown in Figure \ref{fig:EAF09}.

The Ontology database contains inference rules. It takes rules either directly from other SOs as the Primary knowledge, or from a well-established ontology definition, or from its Parameters database.

The Parameters database contains the name-value pairs of data. It takes values either from the IoT devices as the Secondary knowledge, or from its Hypotheses database. Secondary knowledge can become Primary knowledge after the rules in the database has been verified through some inductive learning. A family of machine learning techniques such as rule learning, classification and Bayesian inference can be employed for inductive learning.

The Hypotheses database contains the non-verified inference rules. It takes rules either directly from other SOs as the Invented knowledge, or from any ontology definition, or from its ontology database. Invented knowledge can become Secondary knowledge after some observations that the values have converged to a distribution or a pattern. Statistics can be employed to observe the distribution of data.

The Hypotheses database also learns from the Ontology database. Primary knowledge from the Ontology database can become Invented knowledge after speculative rules are created through abduction techniques. For example, given a rule "apples are red," the speculative rule seeks to conclude if the object is an "apple" when "red" is observed. In this framework, abduction is triggered when an observed value (i.e. red) fails to conclude the class (i.e. apple).

In creating speculative rules, the framework implements two theorems previously discussed. First, it employs Natural Language Processing (NLP) techniques to understand human language. The NLP can find the synonyms, homonyms and categories of a word. In "apples are red," the NLP finds that "red" is a color, and that "apple" is a noun. It therefore seeks to find other nouns, hence creating a rule such as "[noun] are red". The use of NLP in this process implements Theorem 1.

Second, the framework exchanges ontologies with other SOs. Ontologies are structured in the way that the inference rules are understandable by humans. Although SOs are machines, the structure of the ontologies must convey meaning. The Semantic Web ontology illustrated in Figure \ref{fig:EAF03} serves this purpose, as the properties it describes reflect the model in the real world. Hence, it implements Theorem 2.

\subsection{Case Study}
\label{sec:caseStudy}
This section provides an example how the framework can be implemented. It takes the ontology and parameters from published articles and adapts these so that SOs can autonomously build knowledge using the proposed framework.

\subsubsection{Initialization}
\begin{figure*}
	\centering
	\includegraphics[scale=0.6]{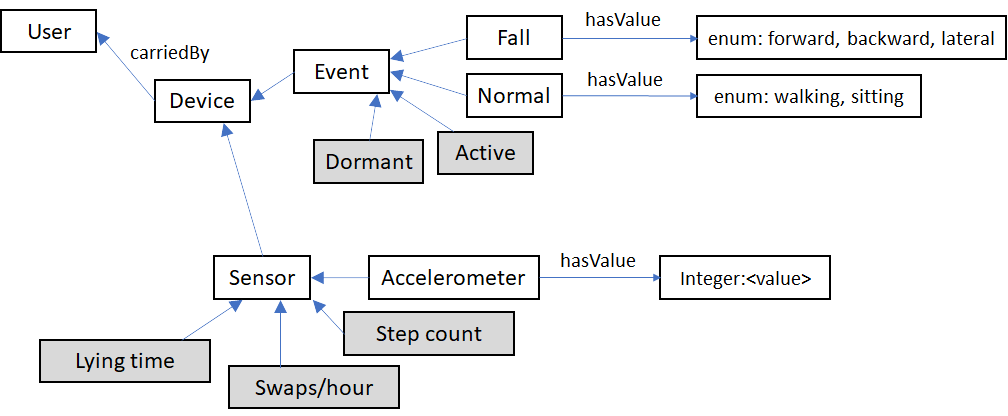}
	\caption{Illustration of entity relationship network of SO1 Ontology database. The shaded ones are learned from SO2.}
	\label{fig:EAF12}
\end{figure*}

A recent study proposed to detect an event when a person falls, for the purpose of sending alerts for support in a smart home domain \cite{EA021}. The study classifies normal daily activities from abnormal activities such as when the person falls. The ontology was described by the non-shaded entities in Figure \ref{fig:EAF12}. In this case study, a Smart Object hereby would be named SO1, and its ontology database is populated accordingly from the fall detection study. Its parameters database is populated with the accelerometer values representing the forward/backward/lateral fall and the name-value pairs representing normal events such as walking and sitting. Its hypotheses database is still empty.

\begin{figure*}
	\centering
	\includegraphics[scale=0.6]{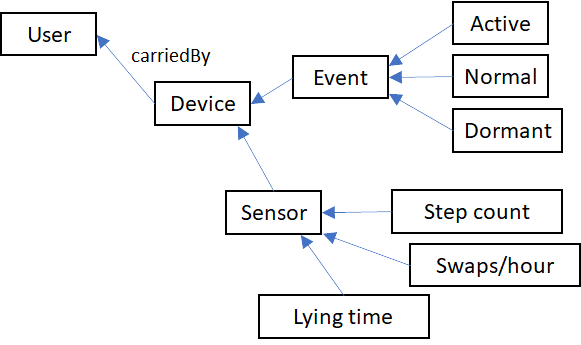}
	\caption{Illustration of entity relationship network of SO2 Ontology database.}
	\label{fig:EAF13}
\end{figure*}

Another study proposed to detect lameness symptoms in dairy cows in the farming domain \cite{EA131} because such a condition is related to milk production’s effectiveness. It places IoT devices on cows to detect lying time, swaps per hour, and step count to derive whether the cows were normal, dormant, or active. Adapting this case to the framework of machine knowledge, the ontology database would have the relationships as shown in Figure \ref{fig:EAF13}. The parameters database would contain the name-value pairs of the lying time, swaps per hour and step count. The hypotheses database would be empty. The system would be named as SO2 in this case.

\subsubsection{Exchange Primary knowledge with other SO}
SO1 learned to expand its knowledge from SO2. It sends a broadcast query to indicate whether other SOs had any Primary knowledge of Event and Sensor or not. SO2 replies by describing its Primary knowledge, i.e. \{Normal, Active, Dormant\} element Event; and \{Lying time, Swaps/hour, Step count\} element Sensor. At this point, SO1 Ontology database is illustrated as shown in Figure 13, having both the shaded and non-shaded entities together.

Since SO1 adopted such Primary knowledge, SO1 queried SO2 to send its Secondary knowledge. Hence, SO2 sends the name-value pairs from the Parameters database. However, the values learned might not be instantly adaptable to SO1 because these values were learned from cows. In contrast, human lying time and step count would have different values to reflect their normal, active, or dormant condition. Similarly, swaps/hour might be a relevant observation for cows but not human.

\subsubsection{From Secondary knowledge becoming Primary knowledge}
The framework guides how to accept or reject the knowledge learned. When Secondary knowledge is generalized through induction, it becomes Primary knowledge. In this case, if it could be verified that SO1 lying time values reflects an SO1 Event (normal, active, dormant, fall), then lying time becomes accepted as an element of the SO1 Sensor. Suppose that after some time, SO1 has collected enough data and updated its lying time values from humans, it will find a range of values that were used to classify some events. Therefore, these lying time values will be accepted as the Secondary knowledge and lying time as an element of Sensor is reflected in the Ontology database. In contrast, swaps per hour are rejected from the Ontology database because its values failed to classify Event.

\subsubsection{From Primary knowledge becoming Invented knowledge}
Rejecting an entity would trigger the framework to perform abduction inferences. It exhaustively tests all rules in the Ontology database. One path is given here as an example: swaps per hour was measured by a sensor; a sensor was a unit of a device; a device was carried by a user. Through abductive reasoning, SO1 speculates that its ”user” was a ”person” (rather than a cow). 

To have the above reasoning outcome, SO1 went through two processes. First, it employed NLP to find the synonyms of the word ”user”, which resulted in many new words such as ”customer”, ”client”, ”patron”, ”prospect”, and ”patient”.

\begin{figure*}
	\centering
	\includegraphics[scale=0.6]{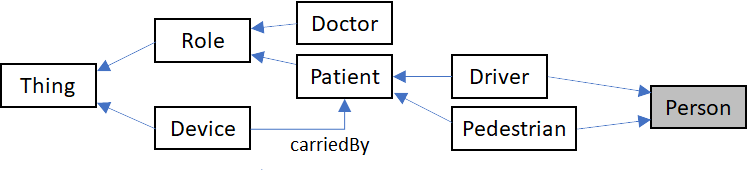}
	\caption{The ontology learned to know what relationships a "patient" has.}
	\label{fig:EAF14}
\end{figure*}

Second, SO1 exchanged Primary knowledge with other SOs to get the ontology structure from the new words. It performed the steps as what previously described to expand knowledge from other SOs, and obtained the relationships as illustrated in Figure \ref{fig:EAF14} (the shaded entity was adapted from a study to monitor traffic in the smart city domain \cite{EA122} and the network (formed by the non-shaded entities) was adapted from a study in the health domain \cite{EA121}).

Thus, the Ontology database had the following relationships: a Device is carried by a User; a Device is carried by a Patient; a Driver is a Patient; a Driver is a Person; a User is synonymous to a Person. Through abduction inference, SO1 speculated that SO1 User is a Person; and a Person is an element of a Role. This relationship was committed in the Hypotheses database.

\begin{figure*}
	\centering
	\includegraphics[scale=0.6]{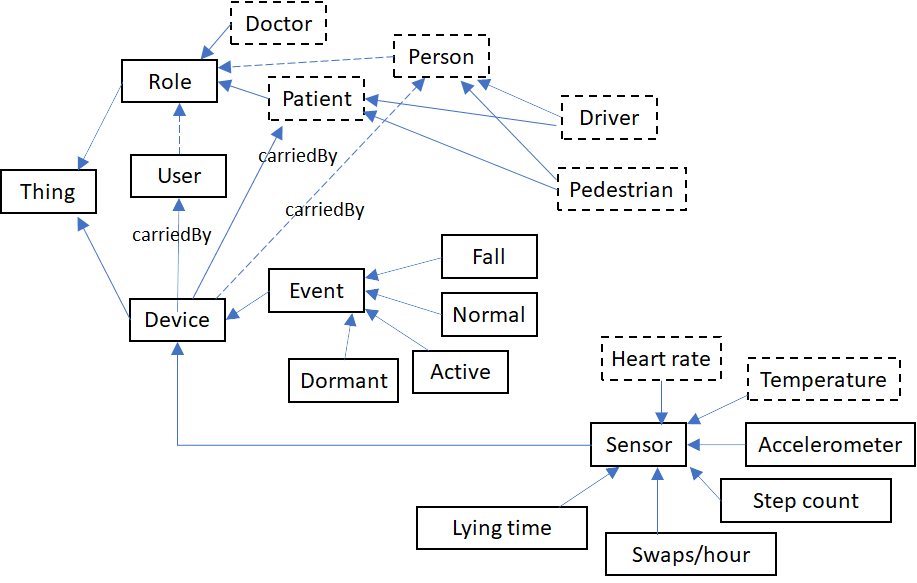}
	\caption{From the attributes of the Sensor, SO1 knows that a User is like a Person.}
	\label{fig:EAF15}
\end{figure*}

Hence, SO1 knowledge would be described as shown in Figure \ref{fig:EAF15}. The dashed entities were added through the above procedure; and the dashed arrows represent the abduction inferences in the Hypotheses database. There would be more inferences drawn which would create a meshed network, but the Figure illustrates only what is relevant for the following discussion.

\subsubsection{From Invented knowledge becoming Secondary knowledge}
In scientific methods, knowledge is gained through observations. When statistics show an acceptable confidence interval from the samples then the hypothesis can be accepted. Therefore, to trigger the statistical analysis, SO1 must have enough data samples. To have enough samples, SO1 should be adopted by the various IoT applications to collect relevant values.

Suppose that SO1 was adopted by a study \cite{EA132} to predict personal thermal comfort based on heart rate and body temperatures obtained from IoT sensors. The human programmer adopted the SO1 ontology described in Figure \ref{fig:EAF15} and created a relationship that "a Device is attached to a Pedestrian" and "Comfortable is an Event". Upon deployment, data samples were collected from the person’s heart rate and body temperature. This showed that the following path had been activated (only one path shown as an example): Heart rate is a unit of a Sensor; a Sensor is a unit of a Device; a Device is carried by a Person; and a Pedestrian is a Person. From this application, there was a sample indicating that "a Pedestrian is a Person".

Interestingly, heart rate, location, and acceleration values were the components used by another study \cite{EA133} to send tailored advertising messages. The human programmer then adopted the solution from the study and added a relationship that "Location is a unit of a Sensor". Hence, this application activated two different paths to name its user:

\begin{itemize}
	\item Heart rate $\rightarrow$ Sensor $\rightarrow$ Device $\rightarrow$ Person (learned from the thermal comfort study \cite{EA132}).
	\item Accelerometer $\rightarrow$ Sensor $\rightarrow$ Device $\rightarrow$ User (learned from the smart home study \cite{EA021}).
\end{itemize}

Once SO1 had collected data samples whose values explained the other relevant relationships (e.g. fall, normal, comfort), then the relationships that were activated were asserted. In this example, the fact that the application indifferently named "Person" and "User" to accomplish a task asserts what earlier was speculative that its User was a Person. 

\subsection{Discussion}
This section demonstrates how the proposed intelligent framework can be applied to various IoT domains (i.e. smart home, farming, smart city, and health) to autonomously exchange and build knowledge. However, the proposed framework does not address how IoT devices can adaptively learn in constrained environments where computing and network resources are very limited (as described by Model 1 in Figure \ref{fig:EAF02}). How better computing and network performance can be achieved is not the focus of our study; rather, we showed how adaptive learning can be achieved within the current IoT models. Hence, the framework leverages the fact that learning, i.e. transforming data to knowledge, can be implemented in the cloud (as in Model 2) or use a high-performance computing solution at the edge (as in Model 3). 

\subsubsection{Latency}
\label{sec:latency}
By adopting Model 2 and 3 as in Figure \ref{fig:EAF02}, the framework can use ontologies in  integrating IoT devices with intelligent systems. Selecting either model would involve assessing how critical the application is. For example, in considering the proper model for a remote, mission-critical application, one can adopt Model 2 (i.e. having the resources in the cloud) provided that the network latency and network availability requirements have been optimized for the remote application. The advantage of this model is that the SO in the cloud can learn from remote SOs in the background, regardless of the network latency connecting to the remote application. The disadvantage is that the network latency can become the bottleneck when the need for real-time decisions increases. As another consideration, one can adopt Model 3 (i.e. having all resources close to the consumer). In this model, the SO can deliver real-time knowledge to the consumer, but it relies on the capacity and availability of the communications network to initiate knowledge exchange with remote SOs.

\subsubsection{Natural Language Processing for machine-to-machine communications}
The framework addresses the current gap in reusing ontologies in various domains by mandating the use of human language in naming the entities and relationships. The branch of AI, Natural Language Processing (NLP), is proposed to intelligently adapt different meaning represented by the ontologies. Hence, in addition to the abovementioned, the contribution of this section is to expand the use of NLP in machine-to-machine communications.

Perhaps the closest parallel with our work is the one described in \cite{EA411}, wherein machine databases gain knowledge from human through NLP parsing, converting semantic representations into logical syntax. The work described the consumers as both human and machines, that build knowledge to and retrieve information from the knowledgebase. Similarly, the work in \cite{EA412} proposed NLP-parsed messages to build a knowledgebase, allowing  machines to build knowledge from human language. Our work differs from these work in that it proposed three databases in the system, i.e. the Ontology, Parameters, and Hypotheses database that allow machine-to-machine communications. Furthermore, our work demonstrated a novel case where machines can autonomously expand their knowledge through the scientific, abduction method.

\subsubsection{Drawbacks}
Although the framework addresses the current challenge of reusing ontologies, it adopts a weak definition of intelligent systems. The Smart Objects given in the case study (i.e. SO1, SO2) are aware of and respond to the changes in their environment, proactively make a decision, and show sociability with other smart objects as well as help others. In addition to these, strong AI would involve displaying emotion and desire, forming a personal character when exposed to a range of situations. However, there is little discussion on whether IoT systems should display a strong AI character, or whether it may be counterproductive. Hence, the framework is sound in proposing new insights in intelligent IoT systems.

\section{Future research opportunities}

The convergence of IoT and large-scale data analytics has created enormous opportunities. Machine intelligence based on IoT data has merged the cyber and physical worlds together and has improved significantly for investigating the real-world challenges from a cyber-physical perspective. The efficiency and reliability of the processes and systems have improved significantly. Now the system operators have better monitoring and control of their systems and processes, and business intelligence people have better insights in understanding their challenges and making informed decisions. Although the convergence of machine intelligence and IoT has opened up many opportunities, there are a few challenges which have constrained their growth. These future research opportunities will enable the seamless integration of IoT with data analytics. The following section addresses some of the possible future directions.

\subsection{Improved cyber security}
For improved decision making, machine learning algorithms heavily rely on IoT data generated and transmitted from the IoT devices. Within an IoT framework, different IoT layers, e.g., perception layer, transportation layer, and application layer, are vulnerable to cyber-attacks. For example, malicious code injection, node tampering, impersonation, Denial of Service Attacks, Routing Attacks, and Data Transit Attack (man-in-the-middle attack, sniffing) are some of the examples of cyber-attacks in an IoT system model \cite{REF}. To secure the IoT system against these cyber threats, it is important to ensure a proper trust management framework. IoT devices themselves require proper attention because most of the devices lack proper security mechanisms. As highlighted in \cite{REF}, the challenges of ensuring IoT security is different compared with the traditional IT security, as discussed below.

First, users can add security solutions in a traditional IT scenario any time; however, most of the IoT devices lack security solutions and others have built-in security solutions, and most of the devices do not support additional security patches, solutions or updates at a later point of time, post-device production.

Second, because of the low memory and processing constraints, only lightweight algorithms are used. As pointed out in \cite{REF}, it is always challenging to strike the proper balance between lower capability and higher security.

Third, in an IoT environment, a wide variety of devices are used. Due to this heterogeneous nature, the security risk is increased due to the integration of devices from different types, technologies, and vendors.  

Fourth, the application layer of IoT suffers from privacy issues. Data leak and eavesdropping could be potential consequences. Fifth, IoT communication protocols are also vulnerable to cyber-attacks and threats including data transit attacks, routing and DoS Attacks, issues in key management, high computational cost of TLS, and lack of user control for privacy.

Our proposed framework for machine learning and knowledge discovery for IoT is the sixth dimension to address cyber security challenges. Intelligent machines communicating with human language may wrongly interpret messages such as, for example, when interpreting homonyms (i.e. the same words that have different meanings such as a "file" which can mean either data or a queue). Current active research in this area is lexical-semantic analysis \cite{EA407} to tell that "april in paris lyrics" refers to the name of a song, while "vacation april in paris" refers to holiday information of a city. However, in the context of collaborating IoT devices sharing knowledge, a malicious node or smart object can intentionally spread fake news. This is possible because homonyms and the varying lexical-semantics can manipulate the content of a message to deceive others \cite{EA408}. Thus, cyber security concerns expand to include machine-to-machine deception techniques.

Therefore, future IoT devices, communication protocols, technologies and the entire platform require improved security solutions by design. Device manufacturers must provide robust in-built security solutions within the devices. Cryptographic algorithms need to be improved and key management must be  made effective. More research is needed on authentication schemes and on Public Key Infrastructure (PKI), focusing on the various IoT technologies and applications. Lightweight but robust intrusion detection techniques need to be developed at the device level, edge or at the cloud of the infrastructure.

\subsection{Machine learning at the edge}
IoT applications will require faster processing and decision making, which trend would move data processing closer to the consumer \cite{EA401}. Sending data to the cloud requires time and high bandwidth. Therefore, analytics at the edge node of the IoT has opened the door for future opportunities. 

One drive behind for this trend is that the use of sensors proliferates in many areas of life and businesses, such as vehicles, manufacturing and health, creating continuous streams of data \cite{EA401,EA402}. Such large amount of data becomes the raw material for businesses and governments to gain insights and new knowledge by employing machine learning techniques. The motivations include competition or creating better policies set out when data is abundant, engendering the urgency to draw knowledge from the data \cite{EA402}. 

Another drive is the convergence between the IoT and critical infrastructure due to their demand for mission-critical applications \cite{EA403}. As discussed in section \ref{sec:latency}, data is processed closer to the consumer when low latency is critical. The current trend is that trivial applications tend to become critical, or can be repurposed to support critical applications. The authors of \cite{EA404} pointed out that e-commerce users need faster time in listing the content of their shopping carts; urban cameras can be deployed to detect a missing child; and self-driving vehicles and air planes generate a large amount of data and require fast decision. Interestingly, the authors of \cite{EA404} pointed out that, in such situation, collaborative learning between the edge devices will increase because ad-hoc machine-to-machine communications do not depend on the cloud data sharing policies.

The challenge against this trend is that edge devices (e.g. mobile devices) tend to have lower computing capabilities than a cloud data center. On the other hand, machine learning data analytics require high processing power and storage. In this context, our proposed framework for machine learning in the IoT environment can address the challenge. Its future work may comprise the study of a federation of edge devices sharing their knowledge. An edge device that is equipped with the highest computing performance can perform the data processing task. An edge device that has the highest bandwidth can communicate with the cloud to offload their computing tasks \cite{EA413}. In \cite{EA405}, a community of IoT devices with enough resources were tasked with making decision on whether an access permission can be granted, on behalf of a low-resource device. In other words, we can further examine a hybrid model comprising Models 2 and 3 (as Figure \ref{fig:EAF02} illustrates) where a group of edge devices communicate to vote which device is responsible to transform data into knowledge. 

\subsection{Scalability}

IoT technologies have enabled the platform to communicate amongst a large number of connected peers. When critical applications and end users require a big number of interconnections, scalability becomes an issue which needs to be addressed. For example, large amounts of data need to be distributed to several end devices, where the devices simultaneously solve computing challenges. Hence, distributed machine learning algorithms with edge computing is a potential solution. It enables computing decisions to take place at the edge, which is closer to the IoT devices. One work in this context is context-aware processing \cite{EA406}, where scalability was achieved by exchanging knowledge rather than some changes in threshold values (as discussed in section \ref{sec:cep}, traditional IoT devices exchange information as a change of state or time). In context-aware processing, information is exchanged in terms of context, allowing the term "morning" to be flexibly defined as "8 am to 12 pm", or when the road traffic is seen as "130 vehicles/hour". This method enables a scalable solution for a network of collaborating devices because only the required information according to the context is exchanged, rather than communicating a number of possible values.

In regard to the framework for machine learning proposed in our work, we note that advancements in NLP research will foster the performance of machine learning tasks in a distributed system. NLP puts context behind data values, allowing connected IoT devices to collaboratively assign labels to their data, learn knowledge from other devices, thereby representing distributed machine learning to solve computing challenges.

\subsection{Hyper convergence}
Within a hyper convergence setting, storage facilities are shared amongst a large number of distributed nodes and their combined performance helps to recover from the resource sharing problem. This paradigm has been shown as the desired architecture for an IoT data analytics framework \cite{J14}. Our framework for machine learning in the IoT environment opens the door to this pathway. The databases shown in Figure \ref{fig:EAF11} (i.e. the Ontology, Parameters, and Hypotheses database) are designed to share their data with other IoT devices in their community network, and their combined knowledge helps to address cross-domain convergence as discussed in section \ref{sec:caseStudy}. However, as pointed in \cite{J14}, the challenge is that current network systems are designed for specific applications. Nodes in a community network are connected with various physical layer technologies and bandwidths, with some nodes being administratively more powerful (e.g. routing information, access control) than others. How a new node can be seamlessly added to an existing community network in the presence of some faults has not been fully addressed. Furthermore, there is a need to examine how a large (and increasing) amount of data can be shared across large-storage nodes such as data centers. Thus, there is a further need to design intelligent solutions for the hyper convergence setting.

Additionally, the convergence of intelligent IoT devices should not only solve the technical challenges but also address regulatory issues. It is essential to justify how intelligent devices can be beneficial in serving humanity rather than only satisfying breakthrough enthusiasts. An example of a current issue is that distributed system solutions consume a large amount of power, which starts to compete with human basic needs \cite{EA403}. Thus, future directions include more studies on regulating the convergence of intelligent machines.

\section{Conclusion}

The IoT paradigm has become an integral part of our daily lives. However, IoT devices are constrained in computation and communication resources, which are the bottlenecks in the development of adaptive, intelligent solutions employing machine learning techniques. Although advances in technologies and platform enhancements pave the way for a future that comprises rapid IoT proliferation, application deployment, and strong analytics of high volume IoT data, we have argued that integrating intelligent solutions from different domains has been proven to be difficult. In this paper, we have justified that our proposed framework for machine learning and knowledge discovery for IoT pave ways to integrate adaptive learning techniques locally, at the edge, through a fog or in the cloud. Consequently, the power of machine learning can be fully harnessed to provide value and benefits to the consumers of IoT technology, in the larger context of things.

Our proposed framework opens up some new challenges for future work in machine-to-machine communications. In cyber security, research can include the study of influence of malicious machines on other devices, that may lead to system compromise. In distributed systems, machines need to be configured to seamlessly distribute resources (hyper convergence), and allow the scalability of connected IoT devices. When a federation of edge devices converge, the concurrent advances in machine learning technology are thus realizable in resource constrained IoT systems, and the future of IoT-enabled human lives, is thus realized.

\begin{acknowledgements}
We thank the anonymous reviewers for their valuable comments which helped us improve the quality, organization and presentation of this paper.
\end{acknowledgements}

%
%

\bibliographystyle{spbasic}      
\bibliography{bib_IoTAI_v13}   

\begin{thebibliography}{114}
\providecommand{\natexlab}[1]{#1}
\providecommand{\url}[1]{{#1}}
\providecommand{\urlprefix}{URL }
\expandafter\ifx\csname urlstyle\endcsname\relax
  \providecommand{\doi}[1]{DOI~\discretionary{}{}{}#1}\else
  \providecommand{\doi}{DOI~\discretionary{}{}{}\begingroup
  \urlstyle{rm}\Url}\fi
\providecommand{\eprint}[2][]{\url{#2}}

\bibitem[{Cis(2015)}]{Cisco_whitepaper}
 (2015) Fog computing and the internet of things: Extend the cloud to where the
  things are.
  \urlprefix\url{{https://www.cisco.com/c/dam/en_us/solutions/trends/iot/docs/computing-overview.pdf}}

\bibitem[{Eri(2019)}]{EricReport}
 (2019) Internet of things forecast mobility report.
  \urlprefix\url{https://www.ericsson.com/en/mobility-report/internet-of-things-forecast}

\bibitem[{EA3(2020{\natexlab{a}})}]{EA302}
 (2020{\natexlab{a}}) Bigquery: Cloud data warehouse. Google Cloud,
  \urlprefix\url{https://cloud.google.com/bigquery/}

\bibitem[{EA3(2020{\natexlab{b}})}]{EA301}
 (2020{\natexlab{b}}) Coremetrics is part of ibm. IBM,
  \urlprefix\url{https://www.ibm.com/au-en/digital-marketing/coremetrics-software}

\bibitem[{Sas([Accessed: 15-01-2020)}]{Sas_ref}
 ([Accessed: 15-01-2020) Predictive analytics history \& current advances. SAS,
  \urlprefix\url{https://www.sas.com/en_au/insights/analytics/predictive-analytics.html}

\bibitem[{Aazam et~al.(2018)Aazam, Zeadally, and Harras}]{EA413}
Aazam M, Zeadally S, Harras KA (2018) Offloading in fog computing for iot:
  Review, enabling technologies, and research opportunities. Future Generation
  Computer Systems 87:278--289

\bibitem[{Akbar et~al.(2015)Akbar, Carrez, Moessner, Sancho, and Rico}]{EA406}
Akbar A, Carrez F, Moessner K, Sancho J, Rico J (2015) Context-aware stream
  processing for distributed {IoT} applications. In: 2015 IEEE 2nd World Forum
  on Internet of Things (WF-IoT), IEEE, pp 663--668

\bibitem[{{Akbar} et~al.(2017){Akbar}, {Khan}, {Carrez}, and {Moessner}}]{J31}
{Akbar} A, {Khan} A, {Carrez} F, {Moessner} K (2017) Predictive analytics for
  complex iot data streams. IEEE Internet of Things Journal 4(5):1571--1582,
  \doi{10.1109/JIOT.2017.2712672}

\bibitem[{{Al-Ali} et~al.(2017){Al-Ali}, {Zualkernan}, {Rashid}, {Gupta}, and
  {Alikarar}}]{J18}
{Al-Ali} AR, {Zualkernan} IA, {Rashid} M, {Gupta} R, {Alikarar} M (2017) A
  smart home energy management system using {IoT} and big data analytics
  approach. IEEE Transactions on Consumer Electronics 63(4):426--434,
  \doi{10.1109/TCE.2017.015014}

\bibitem[{Al-Fuqaha et~al.(2015)Al-Fuqaha, Guizani, Mohammadi, Aledhari, and
  Ayyash}]{EA019}
Al-Fuqaha A, Guizani M, Mohammadi M, Aledhari M, Ayyash M (2015) Internet of
  things: A survey on enabling technologies, protocols, and applications. IEEE
  communications surveys \& tutorials 17(4):2347--2376

\bibitem[{{Alahakoon} and {Yu}(2016)}]{SG2}
{Alahakoon} D, {Yu} X (2016) Smart electricity meter data intelligence for
  future energy systems: A survey. IEEE Transactions on Industrial Informatics
  12(1):425--436, \doi{10.1109/TII.2015.2414355}

\bibitem[{Anwar et~al.(2015)Anwar, Mahmood, and Tari}]{AA2015201}
Anwar A, Mahmood AN, Tari Z (2015) Identification of vulnerable node clusters
  against false data injection attack in an ami based smart grid. Information
  Systems 53:201 -- 212

\bibitem[{Anwar et~al.(2017)Anwar, Mahmood, and Pickering}]{Adnan_JCSS}
Anwar A, Mahmood AN, Pickering M (2017) Modeling and performance evaluation of
  stealthy false data injection attacks on smart grid in the presence of
  corrupted measurements. Journal of Computer and System Sciences 83(1):58 --
  72, \doi{https://doi.org/10.1016/j.jcss.2016.04.005}

\bibitem[{{Anwar} et~al.(2017){Anwar}, {Mahmood}, and {Tari}}]{AA8010846}
{Anwar} A, {Mahmood} AN, {Tari} Z (2017) Ensuring data integrity of opf module
  and energy database by detecting changes in power flow patterns in smart
  grids. IEEE Transactions on Industrial Informatics 13(6):3299--3311

\bibitem[{Ara and Ara(2017)}]{EA134}
Ara A, Ara A (2017) Case study: Integrating iot, streaming analytics and
  machine learning to improve intelligent diabetes management system. In: 2017
  International Conference on Energy, Communication, Data Analytics and Soft
  Computing (ICECDS), IEEE, pp 3179--3182

\bibitem[{Ashraf et~al.(2018)Ashraf, Hussain, Hussain, and Chang}]{EA203}
Ashraf J, Hussain OK, Hussain FK, Chang EJ (2018) Ontology usage analysis
  framework (ousaf). In: Measuring and Analysing the Use of Ontologies,
  Springer, pp 49--62

\bibitem[{Azhar(2011)}]{EA200}
Azhar S (2011) Building information modeling (bim): Trends, benefits, risks,
  and challenges for the aec industry. Leadership and management in engineering
  11(3):241--252

\bibitem[{Bajer(2017)}]{EA009}
Bajer M (2017) Building an iot data hub with elasticsearch, logstash and
  kibana. In: 2017 5th International Conference on Future Internet of Things
  and Cloud Workshops (FiCloudW), IEEE, pp 63--68

\bibitem[{Berkeley(1881)}]{EA114}
Berkeley G (1881) A treatise concerning the principles of human knowledge. JB
  Lippincott \& Company

\bibitem[{Berners-Lee et~al.(2001)Berners-Lee, Hendler, Lassila et~al.}]{EA006}
Berners-Lee T, Hendler J, Lassila O, et~al. (2001) The semantic web. Scientific
  american 284(5):28--37

\bibitem[{Bonomi et~al.(2012)Bonomi, Milito, Zhu, and Addepalli}]{Fog_cisco}
Bonomi F, Milito R, Zhu J, Addepalli S (2012) Fog computing and its role in the
  internet of things. In: Proceedings of the First Edition of the MCC Workshop
  on Mobile Cloud Computing, ACM, New York, NY, USA, MCC '12, pp 13--16

\bibitem[{Bottaccioli et~al.(2017)Bottaccioli, Aliberti, Ugliotti, Patti,
  Osello, Macii, and Acquaviva}]{EA201}
Bottaccioli L, Aliberti A, Ugliotti F, Patti E, Osello A, Macii E, Acquaviva A
  (2017) Building energy modelling and monitoring by integration of {IoT}
  devices and building information models. In: 2017 IEEE 41st Annual Computer
  Software and Applications Conference (COMPSAC), IEEE, vol~1, pp 914--922

\bibitem[{{Bui} and {Jung}(2019)}]{V8}
{Bui} KN, {Jung} JJ (2019) Aco-based dynamic decision making for connected
  vehicles in iot system. IEEE Transactions on Industrial Informatics
  15(10):5648--5655, \doi{10.1109/TII.2019.2906886}

\bibitem[{Burmeister and Schrader(2018)}]{EA084}
Burmeister D, Schrader A (2018) Runtime generation and delivery of guidance for
  smart object ensembles. In: International Conference on Applied Human Factors
  and Ergonomics, Springer, pp 287--296

\bibitem[{Byabazaire et~al.(2019)Byabazaire, Olariu, Taneja, and Davy}]{EA131}
Byabazaire J, Olariu C, Taneja M, Davy A (2019) Lameness detection as a
  service: application of machine learning to an internet of cattle. In: 2019
  16th IEEE Annual Consumer Communications \& Networking Conference (CCNC),
  IEEE, pp 1--6

\bibitem[{{Cao} et~al.(2019){Cao}, {Wachowicz}, {Renso}, and {Carlini}}]{J2}
{Cao} H, {Wachowicz} M, {Renso} C, {Carlini} E (2019) Analytics everywhere:
  Generating insights from the internet of things. IEEE Access 7:71749--71769,
  \doi{10.1109/ACCESS.2019.2919514}

\bibitem[{Cheung et~al.(2018)Cheung, Lin, and Lin}]{EA202}
Cheung WF, Lin TH, Lin YC (2018) A real-time construction safety monitoring
  system for hazardous gas integrating wireless sensor network and building
  information modeling technologies. Sensors 18(2):436

\bibitem[{{Contreras-Castillo} et~al.(2018){Contreras-Castillo}, {Zeadally},
  and {Guerrero-Ibañez}}]{Sherali7892008}
{Contreras-Castillo} J, {Zeadally} S, {Guerrero-Ibañez} JA (2018) Internet of
  vehicles: Architecture, protocols, and security. IEEE Internet of Things
  Journal 5(5):3701--3709

\bibitem[{Dash et~al.(2019)Dash, Shakyawar, Sharma et~al.}]{dash2019big}
Dash S, Shakyawar SK, Sharma M, et~al. (2019) Big data in healthcare:
  management, analysis and future prospects. J Big Data 6 6(54)

\bibitem[{Deligiannis et~al.(2019)Deligiannis, Koutroubinas, and
  Koronias}]{EA033}
Deligiannis P, Koutroubinas S, Koronias G (2019) Predicting energy consumption
  through machine learning using a smart-metering architecture. IEEE Potentials
  38(2):29--34

\bibitem[{Dermeval et~al.(2016)Dermeval, Vilela, Bittencourt, Castro, Isotani,
  Brito, and Silva}]{EA205}
Dermeval D, Vilela J, Bittencourt II, Castro J, Isotani S, Brito P, Silva A
  (2016) Applications of ontologies in requirements engineering: a systematic
  review of the literature. Requirements Engineering 21(4):405--437

\bibitem[{Descartes(2013)}]{EA112}
Descartes R (2013) Ren{\'e} Descartes: Meditations on first philosophy: With
  selections from the objections and replies. Cambridge University Press

\bibitem[{Dey et~al.(2016)Dey, Ling, Syed, Zheng, Landowski, Anderson, Stuart,
  and Tolentino}]{EA142}
Dey A, Ling X, Syed A, Zheng Y, Landowski B, Anderson D, Stuart K, Tolentino ME
  (2016) Namatad: Inferring occupancy from building sensors using machine
  learning. In: 2016 IEEE 3rd World Forum on Internet of Things (WF-IoT), IEEE,
  pp 478--483

\bibitem[{Djuedja et~al.(2019)Djuedja, Karray, Foguem, Magniont, and
  Abanda}]{EA204}
Djuedja JFT, Karray MH, Foguem BK, Magniont C, Abanda FH (2019)
  Interoperability challenges in building information modelling (bim). In:
  Enterprise Interoperability VIII, Springer, pp 275--282

\bibitem[{Ed-daoudy and Maalmi(2019)}]{Ed-daoudy2019}
Ed-daoudy A, Maalmi K (2019) A new internet of things architecture for
  real-time prediction of various diseases using machine learning on big data
  environment. Journal of Big Data 6(1):104

\bibitem[{{Elijah} et~al.(2018){Elijah}, {Rahman}, {Orikumhi}, {Leow}, and
  {Hindia}}]{Agri1}
{Elijah} O, {Rahman} TA, {Orikumhi} I, {Leow} CY, {Hindia} MN (2018) An
  overview of internet of things (iot) and data analytics in agriculture:
  Benefits and challenges. IEEE Internet of Things Journal 5(5):3758--3773,
  \doi{10.1109/JIOT.2018.2844296}

\bibitem[{Endler et~al.(2017)Endler, Briot, e~Silva, de~Almeida, and
  Haeusler}]{EA044}
Endler M, Briot JP, e~Silva FS, de~Almeida VP, Haeusler EH (2017) Towards
  stream-based reasoning and machine learning for iot applications. In: 2017
  Intelligent Systems Conference (IntelliSys), IEEE, pp 202--209

\bibitem[{Fadlullah et~al.(2018)Fadlullah, Pathan, and Gacanin}]{EA137}
Fadlullah ZM, Pathan ASK, Gacanin H (2018) On delay-sensitive healthcare data
  analytics at the network edge based on deep learning. In: 2018 14th
  International Wireless Communications \& Mobile Computing Conference (IWCMC),
  IEEE, pp 388--393

\bibitem[{{Farooq} et~al.(2019){Farooq}, {Riaz}, {Abid}, {Abid}, and
  {Naeem}}]{Agri2}
{Farooq} MS, {Riaz} S, {Abid} A, {Abid} K, {Naeem} MA (2019) A survey on the
  role of iot in agriculture for the implementation of smart farming. IEEE
  Access 7:156237--156271

\bibitem[{{Firouzi} et~al.(2018){Firouzi}, {Farahani}, {Ibrahim}, and
  {Chakrabarty}}]{HC_2}
{Firouzi} F, {Farahani} B, {Ibrahim} M, {Chakrabarty} K (2018) Keynote paper:
  From eda to iot ehealth: Promises, challenges, and solutions. IEEE
  Transactions on Computer-Aided Design of Integrated Circuits and Systems
  37(12):2965--2978, \doi{10.1109/TCAD.2018.2801227}

\bibitem[{Flouris et~al.(2017)Flouris, Giatrakos, Deligiannakis, Garofalakis,
  Kamp, and Mock}]{EA042}
Flouris I, Giatrakos N, Deligiannakis A, Garofalakis M, Kamp M, Mock M (2017)
  Issues in complex event processing: Status and prospects in the big data era.
  Journal of Systems and Software 127:217--236

\bibitem[{Fortino et~al.(2012)Fortino, Guerrieri, and Russo}]{EA083}
Fortino G, Guerrieri A, Russo W (2012) Agent-oriented smart objects
  development. In: Proceedings of the 2012 IEEE 16th international conference
  on computer supported cooperative work in design (CSCWD), IEEE, pp 907--912

\bibitem[{Fortino et~al.(2015)Fortino, Guerrieri, Russo, and Savaglio}]{EA081}
Fortino G, Guerrieri A, Russo W, Savaglio C (2015) Towards a development
  methodology for smart object-oriented iot systems: A metamodel approach. In:
  2015 IEEE international conference on systems, man, and cybernetics, IEEE, pp
  1297--1302

\bibitem[{Garc{\'\i}a-Magari{\~n}o et~al.(2017)Garc{\'\i}a-Magari{\~n}o,
  Lacuesta, and Lloret}]{HC_6}
Garc{\'\i}a-Magari{\~n}o I, Lacuesta R, Lloret J (2017) Agent-based simulation
  of smart beds with internet-of-things for exploring big data analytics. IEEE
  Access 6:366--379

\bibitem[{Gonzalez-Mendoza et~al.(2017)Gonzalez-Mendoza, Velasco-Bermeo, and
  Orozco}]{EA122}
Gonzalez-Mendoza M, Velasco-Bermeo N, Orozco OJL (2017) The traffic status and
  pollutant status ontologies for the smart city domain. In: Mexican
  International Conference on Artificial Intelligence, Springer, pp 95--101

\bibitem[{Granados et~al.(2019)Granados, Chu, Zou, and Zheng}]{EA027}
Granados J, Chu H, Zou Z, Zheng LR (2019) Towards workload-balanced, live deep
  learning analytics for confidentiality-aware iot medical platforms. In: 2019
  IEEE International Conference on Artificial Intelligence Circuits and Systems
  (AICAS), IEEE, pp 62--66

\bibitem[{Griffiths and Ooi(2018)}]{EA401}
Griffiths F, Ooi M (2018) The fourth industrial revolution-industry 4.0 and
  {IoT} [trends in future i\&m]. IEEE Instrumentation \& Measurement Magazine
  21(6):29--43

\bibitem[{Gunduz and Das(2020)}]{GUNDUZ2020107094}
Gunduz MZ, Das R (2020) Cyber-security on smart grid: Threats and potential
  solutions. Computer Networks 169:107094

\bibitem[{Hassanalieragh et~al.(2015)Hassanalieragh, Page, Soyata, Sharma,
  Aktas, Mateos, Kantarci, and Andreescu}]{EA004}
Hassanalieragh M, Page A, Soyata T, Sharma G, Aktas M, Mateos G, Kantarci B,
  Andreescu S (2015) Health monitoring and management using internet-of-things
  (iot) sensing with cloud-based processing: Opportunities and challenges. In:
  2015 IEEE International Conference on Services Computing, IEEE, pp 285--292

\bibitem[{{He} et~al.(2014){He}, {Yan}, and {Xu}}]{V3}
{He} W, {Yan} G, {Xu} LD (2014) Developing vehicular data cloud services in the
  iot environment. IEEE Transactions on Industrial Informatics
  10(2):1587--1595, \doi{10.1109/TII.2014.2299233}

\bibitem[{{He} et~al.(2017){He}, {Mendis}, and {Wei}}]{He7926429}
{He} Y, {Mendis} GJ, {Wei} J (2017) Real-time detection of false data injection
  attacks in smart grid: A deep learning-based intelligent mechanism. IEEE
  Transactions on Smart Grid 8(5):2505--2516

\bibitem[{{Hossain} et~al.(2019){Hossain}, {Khan}, {Un-Noor}, {Sikander}, and
  {Sunny}}]{SG4}
{Hossain} E, {Khan} I, {Un-Noor} F, {Sikander} SS, {Sunny} MSH (2019)
  Application of big data and machine learning in smart grid, and associated
  security concerns: A review. IEEE Access 7:13960--13988,
  \doi{10.1109/ACCESS.2019.2894819}

\bibitem[{{Hossain} and {Muhammad}(2018)}]{HC_7}
{Hossain} MS, {Muhammad} G (2018) Emotion-aware connected healthcare big data
  towards 5g. IEEE Internet of Things Journal 5(4):2399--2406,
  \doi{10.1109/JIOT.2017.2772959}

\bibitem[{Hua et~al.(2015)Hua, Wang, Wang, Zheng, and Zhou}]{EA407}
Hua W, Wang Z, Wang H, Zheng K, Zhou X (2015) Short text understanding through
  lexical-semantic analysis. In: 2015 IEEE 31st International Conference on
  Data Engineering, IEEE, pp 495--506

\bibitem[{Hussein et~al.(2017)Hussein, Bertin, and Frey}]{EA405}
Hussein D, Bertin E, Frey V (2017) A community-driven access control approach
  in distributed {IoT} environments. IEEE Communications Magazine
  55(3):146--153

\bibitem[{Islam(2019)}]{electronics8080898}
Islam SN (2019) A new pricing scheme for intra-microgrid and inter-microgrid
  local energy trading. Electronics 8(8)

\bibitem[{{Islam} et~al.(2019){Islam}, {Baig}, and {Zeadally}}]{SNI8777171}
{Islam} SN, {Baig} Z, {Zeadally} S (2019) Physical layer security for the smart
  grid: Vulnerabilities, threats, and countermeasures. IEEE Transactions on
  Industrial Informatics 15(12):6522--6530

\bibitem[{{Ivanov} et~al.(2015){Ivanov}, {Bhargava}, and
  {Donnelly}}]{agri_external1}
{Ivanov} S, {Bhargava} K, {Donnelly} W (2015) Precision farming: Sensor
  analytics. IEEE Intelligent Systems 30(4):76--80, \doi{10.1109/MIS.2015.67}

\bibitem[{{Jan} et~al.(2019){Jan}, {Ahmed}, {Shakhov}, and {Koo}}]{SG8}
{Jan} SU, {Ahmed} S, {Shakhov} V, {Koo} I (2019) Toward a lightweight intrusion
  detection system for the internet of things. IEEE Access 7:42450--42471,
  \doi{10.1109/ACCESS.2019.2907965}

\bibitem[{Janjua et~al.(2019)Janjua, Vecchio, Antonini, and Antonelli}]{EA028}
Janjua ZH, Vecchio M, Antonini M, Antonelli F (2019) Irese: An intelligent
  rare-event detection system using unsupervised learning on the iot edge.
  Engineering Applications of Artificial Intelligence 84:41--50

\bibitem[{Jeong et~al.(2019)Jeong, Son, and Lee}]{EA024}
Jeong Y, Son S, Lee B (2019) The lightweight autonomous vehicle self-diagnosis
  (lavs) using machine learning based on sensors and multi-protocol iot
  gateway. Sensors 19(11):2534

\bibitem[{{Jiang} et~al.(2019){Jiang}, {Fang}, and {Wang}}]{V1}
{Jiang} T, {Fang} H, {Wang} H (2019) Blockchain-based internet of vehicles:
  Distributed network architecture and performance analysis. IEEE Internet of
  Things Journal 6(3):4640--4649, \doi{10.1109/JIOT.2018.2874398}

\bibitem[{Joshi et~al.(2016)Joshi, Reddy, Reddy, Agarwal, Agarwal, Bagga, and
  Bhargava}]{EA135}
Joshi J, Reddy J, Reddy P, Agarwal A, Agarwal R, Bagga A, Bhargava A (2016)
  Cloud computing based smart garbage monitoring system. In: 2016 3rd
  International Conference on Electronic Design (ICED), IEEE, pp 70--75

\bibitem[{Kant(1781)}]{EA110}
Kant I (1781) Critique of pure reason. Modern Classical Philosophers,
  Cambridge, MA: Houghton Mifflin pp 370--456

\bibitem[{{Kong} et~al.(2018){Kong}, {Xu}, {Cheng}, and {Huang}}]{V6}
{Kong} XTR, {Xu} SX, {Cheng} M, {Huang} GQ (2018) Iot-enabled parking space
  sharing and allocation mechanisms. IEEE Transactions on Automation Science
  and Engineering 15(4):1654--1664, \doi{10.1109/TASE.2017.2785241}

\bibitem[{Krylovskiy(2015)}]{EA022}
Krylovskiy A (2015) Internet of things gateways meet linux containers:
  Performance evaluation and discussion. In: 2015 IEEE 2nd World Forum on
  Internet of Things (WF-IoT), IEEE, pp 222--227

\bibitem[{Laftchiev and Nikovski(2016)}]{EA132}
Laftchiev E, Nikovski D (2016) An iot system to estimate personal thermal
  comfort. In: 2016 IEEE 3rd World Forum on Internet of Things (WF-IoT), IEEE,
  pp 672--677

\bibitem[{Lavassani et~al.(2018)Lavassani, Forsstr{\"o}m, Jennehag, and
  Zhang}]{EA023}
Lavassani M, Forsstr{\"o}m S, Jennehag U, Zhang T (2018) Combining fog
  computing with sensor mote machine learning for industrial iot. Sensors
  18(5):1532

\bibitem[{Li et~al.(2018)Li, Ota, and Dong}]{EA002}
Li H, Ota K, Dong M (2018) Learning iot in edge: Deep learning for the internet
  of things with edge computing. IEEE Network 32(1):96--101

\bibitem[{{Li} et~al.(2019){Li}, {Logenthiran}, {Phan}, and {Woo}}]{SG5}
{Li} W, {Logenthiran} T, {Phan} V, {Woo} WL (2019) A novel smart energy theft
  system (sets) for iot-based smart home. IEEE Internet of Things Journal
  6(3):5531--5539, \doi{10.1109/JIOT.2019.2903281}

\bibitem[{Locke(1841)}]{EA111}
Locke J (1841) An essay concerning human understanding

\bibitem[{Luckham(2011)}]{EA040}
Luckham DC (2011) Event processing for business: organizing the real-time
  enterprise. John Wiley \& Sons

\bibitem[{{Luo} et~al.(2019){Luo}, {Zhang}, {Zhang}, {Yu}, and {Li}}]{V5}
{Luo} X, {Zhang} H, {Zhang} Z, {Yu} Y, {Li} K (2019) A new framework of
  intelligent public transportation system based on the internet of things.
  IEEE Access 7:55290--55304, \doi{10.1109/ACCESS.2019.2913288}

\bibitem[{Matthews(2006)}]{EA113}
Matthews D (2006) Epistemic Humility. In: van Gigch J.P. (eds) Wisdom,
  Knowledge, and Management. C.West Churchman and Related Works Series, vol~2.
  Springer, New York, NY, \doi{https://doi.org/10.1007/978-0-387-36506-0_7}

\bibitem[{Mehdiyev et~al.(2015)Mehdiyev, Krumeich, Enke, Werth, and
  Loos}]{EA046}
Mehdiyev N, Krumeich J, Enke D, Werth D, Loos P (2015) Determination of rule
  patterns in complex event processing using machine learning techniques.
  Procedia Computer Science 61:395--401

\bibitem[{{Merrill}(2010)}]{agri_external2}
{Merrill} W (2010) Where is the return on investment in wireless sensor
  networks? IEEE Wireless Communications 17(1):4--6,
  \doi{10.1109/MWC.2010.5416341}

\bibitem[{Naik(2017)}]{EA018}
Naik N (2017) Choice of effective messaging protocols for {IoT} systems: {MQTT,
  CoAP, AMQP and HTTP}. In: 2017 IEEE international systems engineering
  symposium (ISSE), IEEE, pp 1--7

\bibitem[{{Neto} et~al.(2018){Neto}, {Zhao}, {Rodrigues}, {Camboim}, and
  {Braun}}]{V4}
{Neto} AJV, {Zhao} Z, {Rodrigues} JJPC, {Camboim} HB, {Braun} T (2018)
  Fog-based crime-assistance in smart iot transportation system. IEEE Access
  6:11101--11111, \doi{10.1109/ACCESS.2018.2803439}

\bibitem[{N{\'o}brega et~al.(2019)N{\'o}brega, Gon{\c{c}}alves, Pedreiras, and
  Pereira}]{EA029}
N{\'o}brega L, Gon{\c{c}}alves P, Pedreiras P, Pereira J (2019) An iot-based
  solution for intelligent farming. Sensors 19(3):603

\bibitem[{Paw{\l}owicz et~al.(2018)Paw{\l}owicz, Salach, and Trybus}]{EA141}
Paw{\l}owicz B, Salach M, Trybus B (2018) Smart city traffic monitoring system
  based on 5g cellular network, rfid and machine learning. In: KKIO Software
  Engineering Conference, Springer, pp 151--165

\bibitem[{{Priyashman} and {Ismail}(2019)}]{V7}
{Priyashman} V, {Ismail} W (2019) Signal strength and read rate prediction
  modeling using machine learning algorithms for vehicular access control and
  identification. IEEE Sensors Journal 19(4):1400--1411,
  \doi{10.1109/JSEN.2018.2880736}

\bibitem[{{Rahman} et~al.(2018){Rahman}, {Hassanain}, {Rashid}, {Barnes}, and
  {Hossain}}]{HC_4}
{Rahman} MA, {Hassanain} E, {Rashid} MM, {Barnes} SJ, {Hossain} MS (2018)
  Spatial blockchain-based secure mass screening framework for children with
  dyslexia. IEEE Access 6:61876--61885, \doi{10.1109/ACCESS.2018.2875242}

\bibitem[{Ruta et~al.(2019)Ruta, Scioscia, Loseto, Pinto, and
  Di~Sciascio}]{EA045}
Ruta M, Scioscia F, Loseto G, Pinto A, Di~Sciascio E (2019) Machine learning in
  the internet of things: a semantic-enhanced approach. Semantic Web pp 1--22

\bibitem[{Salhi et~al.(2019)Salhi, Silverston, Yamazaki, and Miyoshi}]{EA139}
Salhi L, Silverston T, Yamazaki T, Miyoshi T (2019) Early detection system for
  gas leakage and fire in smart home using machine learning. In: 2019 IEEE
  International Conference on Consumer Electronics (ICCE), IEEE, pp 1--6

\bibitem[{{Sebastian} et~al.(2019){Sebastian}, {Islam}, {Mahmud}, and
  {Oo}}]{Sebastian8909771}
{Sebastian} AJ, {Islam} SN, {Mahmud} A, {Oo} AMT (2019) Optimum local energy
  trading considering priorities in a microgrid. In: 2019 IEEE International
  Conference on Communications, Control, and Computing Technologies for Smart
  Grids (SmartGridComm)

\bibitem[{Sewak and Singh(2016)}]{EA133}
Sewak M, Singh S (2016) Iot and distributed machine learning powered optimal
  state recommender solution. In: 2016 International Conference on Internet of
  Things and Applications (IOTA), IEEE, pp 101--106

\bibitem[{Shadbolt et~al.(2006)Shadbolt, Berners-Lee, and Hall}]{EA007}
Shadbolt N, Berners-Lee T, Hall W (2006) The semantic web revisited. IEEE
  intelligent systems 21(3):96--101

\bibitem[{{Shah} et~al.(2019){Shah}, {Seker}, {Rathore}, {Hameed}, {Ben Yahia},
  and {Draheim}}]{REF}
{Shah} SA, {Seker} DZ, {Rathore} MM, {Hameed} S, {Ben Yahia} S, {Draheim} D
  (2019) Towards disaster resilient smart cities: Can internet of things and
  big data analytics be the game changers? IEEE Access 7:91885--91903

\bibitem[{{Sharma} and {Wang}(2017)}]{J9}
{Sharma} SK, {Wang} X (2017) Live data analytics with collaborative edge and
  cloud processing in wireless iot networks. IEEE Access 5:4621--4635,
  \doi{10.1109/ACCESS.2017.2682640}

\bibitem[{{Shi} et~al.(2016){Shi}, {Cao}, {Zhang}, {Li}, and {Xu}}]{EC_IoTJ}
{Shi} W, {Cao} J, {Zhang} Q, {Li} Y, {Xu} L (2016) Edge computing: Vision and
  challenges. IEEE Internet of Things Journal 3(5):637--646,
  \doi{10.1109/JIOT.2016.2579198}

\bibitem[{Shi et~al.(2016)Shi, Cao, Zhang, Li, and Xu}]{EA404}
Shi W, Cao J, Zhang Q, Li Y, Xu L (2016) Edge computing: Vision and challenges.
  IEEE internet of things journal 3(5):637--646

\bibitem[{{Siryani} et~al.(2017){Siryani}, {Tanju}, and {Eveleigh}}]{SG3}
{Siryani} J, {Tanju} B, {Eveleigh} TJ (2017) A machine learning
  decision-support system improves the internet of things' smart meter
  operations. IEEE Internet of Things Journal 4(4):1056--1066,
  \doi{10.1109/JIOT.2017.2722358}

\bibitem[{Solmaz et~al.(2018)Solmaz, Mutlu, Alankus, K{\i}l{\i}{\c{c}}, Bayram,
  and Horzum}]{EA136}
Solmaz ME, Mutlu AY, Alankus G, K{\i}l{\i}{\c{c}} V, Bayram A, Horzum N (2018)
  Quantifying colorimetric tests using a smartphone app based on machine
  learning classifiers. Sensors and Actuators B: Chemical 255:1967--1973

\bibitem[{{Stellios} et~al.(2018){Stellios}, {Kotzanikolaou}, {Psarakis},
  {Alcaraz}, and {Lopez}}]{cyber_IoT}
{Stellios} I, {Kotzanikolaou} P, {Psarakis} M, {Alcaraz} C, {Lopez} J (2018) A
  survey of iot-enabled cyberattacks: Assessing attack paths to critical
  infrastructures and services. IEEE Communications Surveys Tutorials
  20(4):3453--3495, \doi{10.1109/COMST.2018.2855563}

\bibitem[{Suenbuel et~al.(2019)Suenbuel, Waldinger, Sikka, and
  Richardson}]{EA411}
Suenbuel A, Waldinger R, Sikka V, Richardson K (2019) Systems and methods for
  natural language processing using machine-oriented inference rules. US Patent
  10,515,154

\bibitem[{{Sun} et~al.(2016){Sun}, {Song}, {Jara}, and {Bie}}]{J8}
{Sun} Y, {Song} H, {Jara} AJ, {Bie} R (2016) Internet of things and big data
  analytics for smart and connected communities. IEEE Access 4:766--773,
  \doi{10.1109/ACCESS.2016.2529723}

\bibitem[{{Sundaravadivel} et~al.(2018){Sundaravadivel}, {Kesavan}, {Kesavan},
  {Mohanty}, and {Kougianos}}]{HC_3}
{Sundaravadivel} P, {Kesavan} K, {Kesavan} L, {Mohanty} SP, {Kougianos} E
  (2018) Smart-log: A deep-learning based automated nutrition monitoring system
  in the iot. IEEE Transactions on Consumer Electronics 64(3):390--398,
  \doi{10.1109/TCE.2018.2867802}

\bibitem[{Terry et~al.(2019)Terry, Harriger, Koepf, Jonnalagadda, Webb-Purkis,
  Gainor, and Griffin}]{EA412}
Terry GA, Harriger JD, Koepf W, Jonnalagadda SR, Webb-Purkis WD, Gainor MS,
  Griffin PD (2019) Systems and methods for enhanced natural language
  processing for machine learning conversations. US Patent App. 16/365,668

\bibitem[{Tsai et~al.(2013)Tsai, Lai, Chiang, and Yang}]{EA001}
Tsai CW, Lai CF, Chiang MC, Yang LT (2013) Data mining for internet of things:
  A survey. IEEE Communications Surveys \& Tutorials 16(1):77--97

\bibitem[{Tsikerdekis and Zeadally(2014)}]{EA408}
Tsikerdekis M, Zeadally S (2014) Online deception in social media.
  Communications of the ACM 57(9):72--80

\bibitem[{{Verma} et~al.(2017){Verma}, {Kawamoto}, {Fadlullah}, {Nishiyama},
  and {Kato}}]{J14}
{Verma} S, {Kawamoto} Y, {Fadlullah} ZM, {Nishiyama} H, {Kato} N (2017) A
  survey on network methodologies for real-time analytics of massive iot data
  and open research issues. IEEE Communications Surveys Tutorials
  19(3):1457--1477, \doi{10.1109/COMST.2017.2694469}

\bibitem[{Vu et~al.(2019)Vu, Nguyen, Nguyen, Nguyen, Massacci, and
  Phung}]{ZB001}
Vu DL, Nguyen TK, Nguyen TV, Nguyen TN, Massacci F, Phung PH (2019) Hit4mal:
  Hybrid image transformation for malware classification. Transactions on
  Emerging Telecommunications Technologies p e3789

\bibitem[{Vyas et~al.(2016)Vyas, Bhatt, and Jha}]{EA402}
Vyas DA, Bhatt D, Jha D (2016) {IoT}: trends, challenges and future scope.
  International Journal of Computer Science \& Communication 7(1):186--197

\bibitem[{Wang et~al.(2019)Wang, Wang, Zhang, and Jin}]{WANG201942}
Wang D, Wang X, Zhang Y, Jin L (2019) Detection of power grid disturbances and
  cyber-attacks based on machine learning. Journal of Information Security and
  Applications 46:42 -- 52

\bibitem[{Wooldridge and Jennings(1995)}]{EA089}
Wooldridge M, Jennings NR (1995) Intelligent agents: Theory and practice. The
  knowledge engineering review 10(2):115--152

\bibitem[{Wu et~al.(2011)Wu, Ranasinghe, Sheng, Zeadally, and Yu}]{SheraliRFID}
Wu Y, Ranasinghe DC, Sheng QZ, Zeadally S, Yu J (2011) Rfid enabled
  traceability networks: a survey. Distrib Parallel Databases 29:397--443

\bibitem[{Xie et~al.(2017)Xie, Zeng, Kurachi, Takada, Li, Li, and Li}]{EA025}
Xie G, Zeng G, Kurachi R, Takada H, Li Z, Li R, Li K (2017) Wcrt analysis of
  can messages in gateway-integrated in-vehicle networks. IEEE Transactions on
  Vehicular Technology 66(11):9623--9637

\bibitem[{Yacchirema et~al.(2019)Yacchirema, de~Puga, Palau, and
  Esteve}]{EA021}
Yacchirema D, de~Puga JS, Palau C, Esteve M (2019) Fall detection system for
  elderly people using iot and ensemble machine learning algorithm. Personal
  and Ubiquitous Computing pp 1--17

\bibitem[{Yacchirema et~al.(2018)Yacchirema, Sarabia-J{\'a}come, Palau, and
  Esteve}]{HC_1}
Yacchirema DC, Sarabia-J{\'a}come D, Palau CE, Esteve M (2018) A smart system
  for sleep monitoring by integrating iot with big data analytics. IEEE Access
  6:35988--36001

\bibitem[{Zeadally and Bello(2019)}]{ZEADALLY2019100074}
Zeadally S, Bello O (2019) Harnessing the power of internet of things based
  connectivity to improve healthcare. Internet of Things p 100074

\bibitem[{Zeadally et~al.(2020)Zeadally, Adi, Baig, and Khan}]{EA403}
Zeadally S, Adi E, Baig Z, Khan I (2020) Harnessing artificial intelligence
  capabilities to improve cybersecurity. IEEE Access 8:23817--23837

\bibitem[{Zekveld and Hancke(2018)}]{EA138}
Zekveld M, Hancke GP (2018) Vibration condition monitoring using machine
  learning. In: IECON 2018-44th Annual Conference of the IEEE Industrial
  Electronics Society, IEEE, pp 4742--4747

\bibitem[{Zeshan and Mohamad(2012)}]{EA121}
Zeshan F, Mohamad R (2012) Medical ontology in the dynamic healthcare
  environment. Procedia Computer Science 10:340--348

\bibitem[{{Zhang} et~al.(2018){Zhang}, {Zhang}, {Liu}, and {Guo}}]{V2}
{Zhang} H, {Zhang} Q, {Liu} J, {Guo} H (2018) Fault detection and repairing for
  intelligent connected vehicles based on dynamic bayesian network model. IEEE
  Internet of Things Journal 5(4):2431--2440, \doi{10.1109/JIOT.2018.2844287}

\end{thebibliography}

\end{document}